\documentclass{article}
\usepackage{amssymb}
\usepackage[all,2cell,ps]{xy}
\usepackage{verbatim}
\usepackage{epsfig}
\usepackage[colorlinks=true, pdfstartview=FitV, linkcolor=blue, citecolor=blue, urlcolor=blue]{hyperref}

\newcommand{\edge}[1]{\ar@{-}[#1]}
\newcommand{\dedge}[1]{\ar@2{->}[#1]}
\newcommand{\tedge}[1]{\ar@3{->}[#1]}
\newcommand{\edged}[1]{\ar@2{-}[#1]}
\newcommand{\edget}[1]{\ar@3{-}[#1]}

\newcommand{\lnode}[1]{*+[o][F-]{\hbox{\tiny #1}}}
\def\t{\widetilde}
\def\ov{\overline}
\def\slint{-\!\!\!-\hspace{-14pt}\int}
\def\ds{\displaystyle}

\def\res{\mathop{\mathrm {res}}\limits_}

\makeatletter
\@addtoreset{equation}{section}

\newtheorem{theorem}{Theorem}[section]
\newtheorem{examp}{Example}[section]
\newtheorem{coroll}{Corollary}[section]
\newtheorem{examps}{Examples}[section]

\newtheorem{lemma}{Lemma}[section]
\newtheorem{remark}{Remark}[section]

\newtheorem{remarks}[remark]{Remarks}
\newtheorem{proposition}{Proposition}[section]
\newtheorem{definition}{Definition}[section]

\def\le{\left}
\def\m{\mathop}
 \def\tr{{\rm Tr}}

\def\ri{\right}
\def\br{\begin{remark}\rm\small}
\def\1{{\bf 1}}
\def\er{\end{remark}}
\def\bt{\begin{theorem}\rm}
\def\et{\end{theorem}}
\def\bc{\begin{coroll}\rm}
\def\ec{\end{coroll}}
\def\brs{\begin{remarks}.\\ \rm\small\begin{enumerate}}
\def\ers{\end{enumerate}\end{remarks}}
\def\bx{\begin{examp}\small}
\def\ex{\end{examp}}
\def\bl{\begin{lemma}\small}
\def\el{\end{lemma}}
\def\bxs{\begin{examps}. \rm\begin{enumerate}}
\def\exs{\end{enumerate}\end{examps}}
\def\bd{\begin{definition}}
\def\ed{\end{definition}}
\def\bp{\begin{proposition}\rm}
\def\ep{\end{proposition}}
\def\be{\begin{equation}}
\def\ee{\end{equation}}
\def\bes{$$}
\def\ees{$$}
\def\bea{\begin{eqnarray}}
\def\eea{\end{eqnarray}}
\def\beas{\begin{eqnarray*}}
\def\eeas{\end{eqnarray*}}
\def \pa{\partial}
\def\C{{\mathbb C}}

\def\Y{{\mathbf  Y}}
\def\X{{\mathbf X}}

\def\wt{\widetilde}
\begin{document}
                                                                  
\fontfamily{cmss}
\fontsize{11pt}{15pt}
\selectfont

\vspace{0.2cm}
\begin{center}
\begin{Large}
\textbf{Two--matrix
  model with semiclassical potentials and extended Whitham hierarchy\footnote{
Work supported in part by the Natural Sciences and Engineering Research Council
of Canada (NSERC), Grant. no. 261229-03.}}
\end{Large}\\
\vspace{1.0cm}
\begin{large} {M.
Bertola}$^{\dagger\ddagger}$\footnote{bertola@mathstat.concordia.ca}
\end{large}
\\
\bigskip
\begin{small}
$^{\dagger}$ {\em Department of Mathematics and
Statistics, Concordia University\\ 7141 Sherbrooke W., Montr\'eal, Qu\'ebec,
Canada H4B 1R6} \\
\smallskip
$^{\ddagger}$ {\em Centre de recherches math\'ematiques,
Universit\'e de Montr\'eal\\ C.~P.~6128, succ. centre ville, Montr\'eal,
Qu\'ebec, Canada H3C 3J7} \\
\end{small}
\bigskip
\bigskip
{\bf Abstract}
\end{center}
\begin{center}
\begin{small}
\parbox{13cm}{
We consider the two-matrix model with potentials whose derivative are
arbitrary rational function of fixed pole structure and the support of
the spectra of the matrices are union of intervals (hard-edges). We derive an
explicit formula for the planar limit of the free energy and we derive
a calculus which allows to compute derivatives of arbitrarily high order by extending classical Rauch's variational formul\ae. The four-points correlation functions are explicitly
worked out. The formalism extends naturally to the computation of {\em
  residue formul\ae} for the tau function of the so-called universal
Whitham hierarchy studied mainly by I. Krichever: our setting extends that moduli space in that there are certain extra data.
}
\end{small}
\end{center}
\tableofcontents

%
%
%
%
%
%
%
%
%
\section{Introduction}
We consider a matrix model
consisting of pairs of Hermitean matrices of size $N$ with an (unnormalized)
probability density of the form
\bea
{\rm d} \mu(M_1,M_2) = {\rm d}M_1 {\rm d}M_2 \exp\le[-\frac 1 \hbar
  \tr(V_1(M_1)+V_2(M_2)-M_1M_2)\ri]\ ;\\
\mathcal Z_N(V_1,V_2,t):= \int {\rm d}\mu\ ,\ \ t:=N\hbar\ .
\eea
Here the {\em potentials} $V_1,V_2$ are required to have rational
derivative; more explicitly we will set
\bea
V_1(x)&:=& V_{1,\infty}(x) + \sum_{\alpha} V_{1,\alpha}(x)\cr
&& V_{1,\infty}(x):= \sum_{K=1}^{d_1+1} \frac {u_{K,\infty}}{K} x^K\
;\qquad V_{1,\alpha}(x):= \sum_{K=1}^{d_{1,\alpha}} \frac
{u_{K,\alpha}}{K(x-Q_\alpha)^K} -u_{0,\alpha}\ln(x-Q_\alpha)\\
V_2(y)&:=& V_{2,\infty}(y) + \sum_{\alpha} V_{2,\alpha}(y)\cr
&& V_{2,\infty}(y):= \sum_{J=1}^{d_2+1} \frac {v_{J,\infty}}{J} y^J\
;\qquad V_{2,\alpha}(y):= \sum_{J=1}^{d_{2,\alpha}} \frac
{v_{J,\alpha}}{J(y-P_\alpha)^J} -v_{0,\alpha}\ln(y-P_\alpha)
\eea
The logarithmic terms in the measure correspond to powers of
determinants. Formally the model is well defined for arbitrary
potentials with complex coefficients, provided that we
constrain the spectrum to belong to certain contours in the complex
plane along the lines explained in \cite{BEH}. In this case, however,
the matrices $M_i$ are no longer Hermitean but only normal
(i.e. commuting with their Hermitean-adjoint).\\
If we insist on a {\em bona fide} Hermitean model we should impose
that $V_i$ are real functions, bounded from below on the real axis.\\
In addition to these data we impose that the spectrum contains
segments with extrema $\{X_i\}$ for the first matrix and $\{Y_j\}$ for the
second matrix ({\bf hard--edges} of the spectra): in the case of Hermitean matrices then we would be restricting the support of the spectra to some arbitrary union of intervals.

It is used as a working hypotheses that the following limit exists
\be
\mathcal F (V_1,V_2,t):= \lim_{N\to\infty} \frac 1 {N^2} \ln \mathcal
Z_N\ ,
\ee
where $t=N\hbar$ is kept fixed in  the limit process.\\
 This model has been analyzed in two papers \cite{eynardsemi} and \cite{bertosemi} from two opposite points of view: in \cite{eynardsemi} were derived the formal properties of the spectral curve and the loop equations in the large $N$ limit, whereas in \cite{bertosemi} were considered the properties of the associated biorthogonal polynomials and the differentials equations they satisfy for finite $N$, together with certain Riemann--Hilbert data.

 The loop-equations show that in the
planar limit the resolvents of the two matrices
\be
W(x) = \lim_{N\to\infty} \frac 1 N \le< \frac 1 {x-M_1}\ri>\ ,\ \
\wt W(y) = \lim_{N\to\infty} \frac 1 N \le< \frac 1 {y-M_2}\ri>
\ee
satisfy an {\em algebraic} equation if we replace $y=Y(x):=
W(x)-V'_1(x)$. This means that there is a rational expression that
defines a (singular) curve in $\Sigma \hookrightarrow \mathbb
P^1\times \mathbb P^1$ --hereby referred to as {\bf spectral curve}--
\be
E(x,y)=0\label{algeq}.
\ee
and that the cuts of the branched covers $x:\Sigma \to \mathbb P^1$
and $y:\Sigma \to \mathbb P^1$ describe the support of the asymptotic density of
eigenvalues, and the jumps across these cuts describe the densities
themselves.

From the finite $N$ analysis the spectral curve \cite{BEH} arises naturally in conjunction with the  ODE satisfied by the associated biorthogonal polynomials; indeed any $s_2$ consecutive biorthogonal polynomials (were $s_2$ is the total degree of the rational function $V'_2(y)$) satisfy a $(s_2+1)$ system of first order ODEs, namely an equation of the form 
\be
\pa_x \Psi_N(x) = D_{N}(x) \Psi_N
\ee
and the spectral curve is nothing but $E_N(x,y)   = det(y\1 - D_N(x)) =0 $.
While clearly certain properties are valid only for finite $N$ or in the infinite limit, certain other properties can be read off both regimes: for instance it can be seen \cite{bertosemi} that at the hard--edges the matrix $D_N(x)$ has simple poles with nilpotent rank-one residue. This implies  certain local  structure of the spectral curve $y(x)$ above these points.

In an algebro-geometric approach the functions $x,y$ themselves are
meromorphic functions on the spectral curve $\Sigma$ with 
specified pole structure and specified singular part near the poles.
The loop equations also provide a first-order overdetermined set of
compatible equations for the free energy; these however are not sufficient to uniquely
determine the partition function because the polar data of the
functions $x,y$ need to be supplemented by extra parameters. This is a
purely algebro-geometric consideration but they also can be
heuristically justified along the lines of \cite{BDE}. It turns
out that the extra unspecified parameters can be taken as the 
contour-integrals
\be
\epsilon_\gamma := \oint_\gamma y{\rm d}x\ ,
\ee
over a maximal set of ``independent'' non-intersecting contours. The reader with some
background in algebraic geometry will recognize that there are
$g=$genus$(\Sigma)$ such contours\footnote{More appropriately one
  should consider only the imaginary parts of these integrals over the
full homology of the curve.}.
These parameters are often called ``filling fractions'' and in
principle they be should uniquely determined by the potentials; the loop
equations cannot determine the filling fraction but can determine the
variations of the Free energy w.r.t. them. This way one obtains an
extended set of (still compatible) PDEs for $\mathcal F$ in terms of
the full moduli of the algebro-geometric problem: we call this
function the {\bf non-equilibrium free energy}.  In this situation
one can actually integrate the PDEs and provide a formula for the
planar limit, $\mathcal F$. \\
Note that the actual free energy of the model is obtained by
expressing the filling fraction implicitly as functions of the
potentials via the equations
\be
\pa_{\epsilon_j} \mathcal F(V_1,V_2,t,\underline\epsilon) \equiv 0.
\ee
Implicit solution yields $\underline \epsilon = \underline \epsilon
(V_1,V_2,t)$; the resulting function
\be
\mathcal G (V_1,V_2,t):= \mathcal F(V_1,V_2,t,\underline \epsilon
(V_1,V_2,t))\ ,
\ee
will be called the {\bf equilibrium free energy}. The distinction is
important when computing the higher order derivatives of $\mathcal G$,
inasmuch as they differ by the higher order derivatives of $\mathcal
F$ by virtue of the chain-rule; indeed while
\be
\frac {\delta \mathcal G}{\delta V_1(x)} = \frac {\delta \mathcal
  F}{\delta V_1(x)} \bigg|_{\underline \epsilon =\underline \epsilon
(V_1,V_2,t)}\ ,
\ee
(since $\pa_{\epsilon_j}\mathcal F =0$) for the second and higher
variations the equations do differ, for
example
\be
\frac {\delta^2 \mathcal G}{\delta V_1(x)\delta V_1(x')} = \le[\frac {\delta^2 \mathcal
  F}{\delta V_1(x)\delta V_1(x')}  + \sum_{j=1}^g \frac {\delta\pa_{\epsilon_j} \mathcal
  F}{\delta V_1(x)}\frac {\delta \epsilon_j}{\delta V_1(x)}\ri]
 \bigg|_{\underline \epsilon =\underline \epsilon
(V_1,V_2,t)}\
\ee
 We
will provide simple formulas for both $\mathcal G,\mathcal F$.
\par\vskip 6pt
The main approach of this paper is similar to that of \cite{F1, F2, 
  chaineloop}, namely that of ascertain the algebro-geometric data in
a convenient abstract formulation which provides an explicit
formula for $\mathcal F$. Once this step is accomplished we also build
the formalism for the ``calculus'' that allows to compute arbitrarily
high order partial derivatives;  we
recall that the derivatives of $\mathcal F$ represent higher order
correlators of the spectral invariants of the model in this planar
limit and also, the coefficients of their expansion in the parameters of
the potentials can be related to enumerative problems of polyvalent
fat-graph on the sphere. This calculus relies on an extension of Rauch's variational formul\ae\ to higher order variations (usual Rauch's formul\ae\ are used to describe first order variations).\par
We want to mention an interesting byproduct of the formalism developed
for our calculus: indeed, with minor modifications mainly in the
notation, the same calculus can be applied to computing variation of arbitrarily high order of the ``tau'' function of the universal Whitham
hierarchy \cite{Kric, teod} which in turn is of relevance for the
Seiberg--Witten model. This application is developed in the Appendix.

In this application in fact we obtain an {\bf extension} of the Whitham hierarchy, because on of the two  primary differentials $d\X, d\Y$ must have some of its simple poles at the points where the other has some of its simple zeroes: this requirement follows from the structure of the spectral curve above the hard-edge points. In other words in this moduli space the differentials have also some poles at non-marked points on the spectral curve.

\subsection{Bergman kernel}
\label{bergman}
We recall the definition of
 the Bergman kernel\footnote{Our use of the term ``Bergman kernel''
 is slightly unconventional, since more commonly the Bergman kernel
 is a reproducing kernel in the $L^2$ space of holomorphic
 one-forms. The kernel that we here name ``Bergman'' is sometimes
 referred to as the ``fundamental symmetric  bi-differential''. We
 borrow the (ab)use of the name ``Bergman'' from \cite{korotkin}.}
 a classical object in complex geometry which can
be represented in terms of prime forms and Theta functions.
In fact we will not need any such sophistication because we are going
to use only its fundamental properties (that uniquely determine it).
The Bergman kernel $\Omega (\zeta,\zeta')$ (where $\zeta,\zeta'$
denote here and in the following abstract
points on the curve) is a bi-differential on $\Sigma_g\times\Sigma_g$
with the properties
\bea
&&\hbox{Symmetry: }\qquad\Omega(\zeta,\zeta')= \Omega(\zeta',\zeta)\\
&&\hbox{Normalization: }\qquad\oint_{\zeta'\in a_j}\Omega(\zeta,\zeta') = 0 \\
&&\oint_{\zeta'\in b_j} \Omega(\zeta,\zeta') = 2i\pi \omega_j(\zeta) \
= \ \hbox { the holomorphic
  normalized Abelian differential}\ .
\eea
It is holomorphic everywhere on $\Sigma_g\times\Sigma_g\setminus
\Delta$,  and it has a double pole on the diagonal $\Delta:=\{\zeta=\zeta'\}$:
namely,  if $z(\zeta)$ is any coordinate, we have
\be
\Omega(\zeta,\zeta') \m{\simeq}_{\zeta\sim  \zeta'} \le[\frac 1{(z(\zeta)-z(\zeta'))^2}  + \frac
1 6 S_B(\zeta) + \mathcal O(z(\zeta)-z(\zeta'))\ri]{\rm d}z(\zeta){\rm d}z(\zeta')\ ,
\ee
where the very important quantity $S_B(\zeta)$ is the `` Bergman projective
connection'' (it transforms like the Schwartzian derivative under
changes of coordinates).\par
It follows also from the general theory that any normalized Abelian
differential of the third kind with simple poles at two points $z_{-}$
and $z_{+}$ with residues respectively $\pm 1$ is obtained from the
Bergman kernel as
\be
{\rm d}S_{z_+,z_-}(\zeta) = \int_{\zeta'=z_-}^{z_+}
\Omega(\zeta,\zeta')\ .
\ee
For later purposes we introduce the dual Bergman kernel defined by
\be
\wt\Omega(\zeta,\zeta') := \Omega(\zeta,\zeta') - 2\pi i\sum_{j,k=1}^{g}
 \omega_j(\zeta)\omega_k(\zeta') (\mathbb B^{-1})_{jk}\ ,\label{dual}
\ee
where $\mathbb B$ is the matrix of $b$-periods
\be
\mathbb B_{ij} = \mathbb B_{ji} = \oint_{b_j} \omega_i\ .
\ee
In fact $\wt\Omega$ is conceptually no different from $\Omega$, being
 just normalized so that $\oint_{b_j}\wt \Omega\equiv 0$. We keep the
 distinction only for later practical purposes.
\subsubsection{Prime form}
For the sake of completeness  we recall here that the definition of the prime form
$E(\zeta,\zeta')$.
\bd
The prime form $E(\zeta,\zeta')$ is the $(-1/2,-1/2)$ bi-differential on
$\Sigma_g\times \Sigma_g$
\bea
E(\zeta,\zeta') = \frac {\Theta\le[\alpha\atop \beta\ri]
  (\mathfrak u(\zeta)-\mathfrak u(\zeta'))}
{h_{\le[\alpha\atop \beta\ri]} (\zeta) h_{\le[\alpha\atop \beta\ri]}  (\zeta') }
\\
h_{\le[\alpha\atop \beta\ri]} (\zeta)^2 := \sum_{k=1}^{g}
\pa_{\mathfrak u_k}\ln\Theta\le[\alpha\atop \beta\ri]\bigg|_{\mathfrak
  u=0} \omega_k(\zeta)\ ,
\eea
where $\omega_k$ are the normalized Abelian holomorphic differentials,
$\mathfrak u$ is the corresponding Abel map and $\le[\alpha\atop
  \beta\ri]$ is a half--integer odd characteristic (the prime form does
not depend on which one).
\ed
Then the relation with the Bergman kernel is the following
\be
\Omega(\zeta,\zeta') = {\rm d}_\zeta {\rm d}_{\zeta'} \ln
E(\zeta,\zeta') = \sum_{k,j=1}^{g}\pa_{\mathfrak u_k} \pa_{\mathfrak u_j}
\ln\Theta\le[{\alpha \atop \beta}\ri] \bigg|_{\mathfrak u(\zeta)-\mathfrak u
  (\zeta')} \omega_k(\zeta) \omega_j(\zeta')
\ee
\br
In genus zero -of course- there are no theta functions: however there
is a Bergman kernel with the same properties, given simply by (using
the standard coordinate on the complex plane)
\be
\Omega(z,z') = \frac {{\rm d}z{\rm d}z'}{(z-z')^2}\ .
\ee
\er
\section{Setting and notations}
\label{setting}
We extend the setting of the paper \cite{F1}\footnote{The
  functions that there were denoted with $P,Q$ are here denoted with
  $\Y,\X$.}  and we will  work with the following data:
a (smooth) curve $\Sigma_g$ of genus $g$ with $2+K+L$ distinct marked
points $\infty_\X,p_1,\dots,p_K$, $\infty_\Y,q_1,\dots,q_L$ and  two functions
$\X$ and $\Y$ with the following pole structure;
\begin{enumerate}
\item The function $\X$ has the following divisor of poles
\be
(\X)_- = \infty_\X + d_{2,\infty}\infty_\Y + \sum_{\alpha=1}^{H_1}
(d_{2,\alpha}+1) p_\alpha + \sum_{\ell=1}^{K_1} \eta_\ell
\ee
\item The function $\Y$ has the following divisor of poles
\be
(\Y)_- = \infty_\Y + d_{1,\infty}\infty_\X + \sum_{\alpha=1}^{H_2}
(d_{1,\alpha}+1) q_\alpha + \sum_{\ell=1}^{K_2} \xi_\ell
\ee
\item The differential ${\rm d}\X$ vanishes (simply) at the (non-marked) points $\{\xi_\ell\}$ and viceversa the differential ${\rm d}\Y$ vanishes (simply) at the points $\{\eta_\ell\}$.
\end{enumerate}
All the points entering he above formul\ae\ are assumed to be pairwise
distinct.
The points of the pole divisors which are not marked (the $\xi_\ell,
\eta_\ell$) will be  called  ``hard-edge''. As hinted at in the introduction  these requirement  follow from either the loop equations \cite{eynardsemi} or the exact form of the spectral curve \cite{bertosemi}: the points $Q_\alpha:= \X(q_\alpha)$ and $X_j:= \X(\xi_j)$ are the positions of the poles of the derivatives of the potential $V_1'(X)$ and the hard-edges in the $\X$--plane (and conversely for $\Y$): the fact that the ODE for the biorthogonal polynomials  has simple poles with nilpotent, rank--one residue at the points $X_j$, $j=1,\dots$,  implies that the differential ${\rm d}\X$ vanishes at one of the points above $X_j$, at which the eigenvalue $\Y$ has a simple pole.

Under these assumptions we can write the following asymptotic expansions
\bea
\Y &=& \left\{
\begin{array}{ll}
\ds \Y = \sqrt{ \frac {-2 R_j}{\X-X_j}} + \mathcal O(1) & \hbox { near }
\xi_j,\hbox{ (here $X_j:= \X(\xi_j)$)}\\
\ds -\sum_{K=0}^{d_{1,\alpha}} \frac {u_{K,\alpha}}{(\X-Q_\alpha)^{K+1}} +
  \mathcal
  O(1) & \hbox{ near } p_j, \hbox{ (here $Q_\alpha:=\X(q_\alpha)$)}\\
\ds \sum_{K=1}^{d_{1,\infty}+1} u_{K,\infty} \X^{K-1} -\frac {t+\sum_\alpha u_{0,\alpha}} \X +\mathcal O(\X^{-2})&
  \hbox{near $\infty_\X$}
\end{array}\ri.\cr
\X &=& \left\{
\begin{array}{ll}
\ds \X = \sqrt{ \frac {-2 S_j}{\Y-Y_j}} + \mathcal O(1) & \hbox { near }
\eta_j,\hbox{ (here $Y_j:= \Y(\eta_j)$)}\\
\ds -\sum_{J=0}^{d_{2,\alpha}} \frac {v_{J,\alpha}}{(\Y-P_\alpha)^{J+1}} + \mathcal
  O(1) & \hbox{ near
  } p_\alpha, \hbox{ (here $P_\alpha:=\Y(p_\alpha)$)}\\
\ds \sum_{J=1}^{d_{2,\infty}+1}v_{J,\infty}  \Y^{J-1} -\frac {t+\sum_\alpha v_{0,\alpha}}   \Y +\mathcal O(\Y^{-2})&
  \hbox{near $\infty_\Y$}
\end{array}\ri.
\label{coord1}
\eea
The asymptotics above imply immediately that there exist two rational
functions which we denote by $V_1'$ and $V_2'$ such that
\be
\le(\Y-V'_1(\X)+\frac t\X\ri){\rm d}\X\ ,\qquad \le(\X-V'_2(\Y)+\frac t \Y\ri){\rm d}\Y
\ee
are holomorphic differentials in the vicinity of the points
$\{\infty_\X,q_\alpha,\alpha\geq 1\}$ and
$\{\infty_\Y,p_\alpha,\alpha\geq 1\}$, respectively. For later
reference we spell out these  functions
\bea
V_1(x)&:=& V_{1,\infty}(x) +
  \sum_{\alpha} \le(V_{1,\alpha}(x) - u_{0,\alpha} \ln(x-Q_\alpha)\ri)\cr
&& V_{1,\infty}(x):= \sum_{K=1}^{d_1+1} \frac {u_{K,\infty}}{K} x^K\
;\qquad V_{1,\alpha}(x):= \sum_{K=1}^{d_{1,\alpha}} \frac
{u_{K,\alpha}}{K(x-Q_\alpha)^K}\\
V_2(y)&:=& V_{2,\infty}(y)
 +  \sum_{\alpha}\le( V_{2,\alpha}(y)-v_{0,\alpha}\ln(y-P_\alpha)\ri)\cr
&& V_{2,\infty}(y):= \sum_{J=1}^{d_2+1} \frac {v_{J,\infty}}{J} y^J\
;\qquad V_{2,\alpha}(y):= \sum_{J=1}^{d_{2,\alpha}} \frac
{v_{J,\alpha}}{J(y-P_\alpha)^J}
\eea
A local set of coordinates for the moduli space of these data is
provided by the coefficients $\{u_{K,\alpha},v_{J,\alpha}, t: \ \alpha
= \infty,1,2,\dots\}$, the
position of the poles $\{Q_\alpha,P_\alpha\}_{\alpha=1,\dots}$, the position of the
hard-edge divisors $\{X_j,Y_j\}$ together with the so-called
filling fractions
\be
\epsilon_j:= \oint_{a_j} \Y{\rm d}\X.\label{coord2}
\ee
\subsection{Planar limit of the free energy}
The planar limit of the free energy is defined by the following set of compatible
equations
\bea
&\begin{array}{l|l}
\ds  \pa_{u_{K,0}}\mathcal F = U_{K,0} :=   - \frac 1 K \res{\infty_\X} \X^K \Y{\rm d}\X &
\ds \pa_{v_{J,0}}\mathcal F = V_{J,0} :=   - \frac 1 J \res{\infty_\Y}
\Y^J \X{\rm d}\Y \\
\ds 
\pa_{u_{K,\alpha}}\mathcal F = U_{K,\alpha} := - \frac 1 K \res{q_\alpha} \frac 1{(\X-Q_\alpha)^K}\Y{\rm d}\X &
\ds
 \pa_{v_{J,\alpha}}\mathcal F = V_{K,\alpha} := - \frac 1 J \res{p_\alpha} \frac 1{(\Y-P_\alpha)^J} \X{\rm d}\Y \\
\ds 
\pa_{u_{0,\alpha}}\mathcal F = U_{0,\alpha}:= \slint^{\infty_\X}_{q_\alpha}\Y{\rm d}\X 
 &
 \ds
\pa_{v_{0,\alpha}}\mathcal F = V_{0,\alpha}:= \slint^{\infty_\Y}_{p_\alpha}\X{\rm d}\Y  \\
\ds  \pa_{X_j} \mathcal F=R_j := \frac 1 2 \res{\xi_j} \Y^2{\rm d}\X  &
\ds \pa_{Y_j} \mathcal F=S_j := \frac 1 2 \res{\eta_j} \X^2{\rm d}\Y  
\\
\hspace{1cm} \ds\pa_{Q_\alpha} \mathcal F = \res{q_\alpha}
\le(V'_{1,\alpha}(\X) -\frac{u_{0,\alpha}}{(\X-Q_\alpha)} \ri)
 \Y{\rm d}\X &
 \ds\hspace{1cm} \pa_{P_\alpha} \mathcal F = \res{p_\alpha}
\le(V'_{2,\alpha}(\Y) - \frac {v_{0,\alpha}}{(\Y-P_\alpha)}\ri)\X{\rm d}\Y
\end{array} &\cr
& \ds  \pa_t \mathcal F =\mu := 
\slint_{\infty_\Y}^{\infty_\X} \Y{\rm d}\X - \sum_{\alpha\geq 1}v_{0,\alpha} =
\slint_{\infty_\X}^{\infty_\Y} \X{\rm d}\Y - \sum_{\alpha\geq 1}u_{0,\alpha}
&
\cr
&\ds 
 \pa_{\epsilon_j} \mathcal F = \Gamma_j:= \frac 1 {2i\pi}\oint_{b_j}
\Y{\rm d}\X\ .\label{firstders}&
\eea
In these formul\ae\ the symbol $\slint$ stands for the regularized
integral obtained by subtraction of the singular part in the local
parameter as follows: \par
{\bf (i)} at $\infty_\X$ ($\infty_\Y$) the local parameter is $z=\X^{-1}$
($\wt z=\Y^{-1}$);\par
{\bf (ii)} at $q_\alpha$ ($p_\alpha$) the local parameter is $z_\alpha =
\X-Q_\alpha$ ($z_{\wt\alpha}=\Y-P_\alpha$).\\
The regularization is then defined as follows: if $z$ is any of the
above local parameters then
\be
\slint^0 \omega := \lim_{\epsilon\to 0} \int^\epsilon \omega -
f(\epsilon)\ ,
\ee
where $f(z)$ is defined as the antiderivative (without constant) of
the singular part of $\frac \omega{{\rm d}z}$ as a function of $z$
(near $z=0$). For example
\bx
The  regularized integral  according to the definition is
\be
\slint_{\infty_\X}^{q_\alpha} \Y {\rm d}\X := \lim_{\epsilon\to
  q_\alpha}\lim_{R\to \infty_\X} \int_{R}^\epsilon \Y{\rm d}\X +
V_{1,\infty}(\X(R))-\le(t-\sum_\alpha u_{0,\alpha}\ri)\ln(\X(R)) - V_{1,\alpha}(\X(\epsilon)).
\ee
\ex
 The two expressions for $\mu$ in (\ref{firstders}) are proven to be
 equivalent (thus showing the symmetry in the r\^oles of $\X$ and
 $\Y$) by integration by parts, paying attention at
the definition of the regularization (which involves as local
parameters $\X^{-1}$ and $\Y^{-1}$ at the two different poles); indeed we have
\bea
\slint_{p}^{\infty_\X} \Y{\rm d}\X = \lim_{\epsilon\to\infty_\X} \le(
\int_{p}^\epsilon \Y{\rm d}\X  - V_{1,\infty}(\X(\epsilon)) +
(t+\sum_{\alpha} u_{0,\alpha}) \ln \X(\epsilon)  \ri) =\\
= \lim_{\epsilon\to\infty_\X} \bigg( -
\int_{p}^\epsilon \X{\rm d}\Y +\X(\epsilon)
\hspace{-39pt}\mathop{\Y(\epsilon)}^{
  V'_{1\infty}(\X) - \X^{-1}(t+\sum u_{0,\alpha})+\dots\atop ||}\hspace{-39pt}
-\X(p)\Y(p) + V_{1,\infty}(\X(\epsilon)) +
(t+\sum_{\alpha} u_{0,\alpha}) \ln \X(\epsilon)  \bigg) =\\=
-\slint_p^{\infty_\X} \X{\rm d}\Y - \X(p)\Y(p) - (t+\sum_{\alpha} u_{0,\alpha})
\eea
together with a similar formula for the symmetric expression
\be
\slint^{p}_{\infty_\Y} \Y{\rm d}\X = \X(p)\Y(p)+ \le(t+\sum_{\beta}
v_{0,\beta}\ri) - \slint_{\infty_\Y}^p \X{\rm d}\Y\ .
\ee
Combining the two one has
\bea
\mu = \slint_{\infty_\Y}^{\infty_\X} \Y{\rm d}{\X}   -
\sum_{\beta} v_{0,\beta}  = \slint_{\infty_\X}^{\infty_\Y} \X{\rm d}{\Y}  -
\sum_{\alpha} u_{0,\alpha}\ .
\eea

In full generality, given any meromorphic differential and local parameters around its poles one
can give completely explicit formul\ae\ for its regularized integrals (see App. \ref{genreg}).
In our specific setting we give explicit formul\ae\ of the previous regularized integrals 
in terms of canonical differentials of the third kind in
App. \ref{regularizedintegral}.\par\vskip 2pt

\noindent\parbox[b]{10cm}{\phantom{-}\ \ 
We make also the important remark that in order for the above formul\ae\ to
make sense we must perform some surgery on the surface by cutting it
along a choice of  the $a,b$-cycles and by performing some mutually non-intersecting
cuts between the poles
with nonzero residues of the differential $\Y{\rm d}\X$. We achieve
this goal by choosing some segments on the surface joining $\infty_\X$
to $\infty_P$, $\infty_\X$ to $q_\alpha$ and $\infty_\Y$ to
$p_\alpha$. The result of this dissection is a simply connected domain
where $\X$, $\Y$ are meromorphic functions and where the
regularizations involving logarithms are defined by taking the
principal determination.\\
\phantom{-}\ \ The compatibility of  equations (\ref{firstders}) for $\mathcal F$ can be shown
 by taking the cross-derivatives. We now  briefly recall, for the reader's sake, how to compute them since much of the formalism is needed in the following.
The main tool is the previously defined  Bergman kernel
(Sect. \ref{bergman})  providing an effective way of
writing formulas for first, second and third-kind normalized
differentials on the Riemann-surface. This is needed when computing
the cross derivatives of the free energy since the differentials $\pa
\Y{\rm d}\X$ and $\pa \X{\rm d}\Y$ (here $\pa$ is any variation of the
coordinates) can be identified with certain canonical differentials.
}\,\,
\parbox[b]{8.1cm}{
\epsfxsize 8cm
\epsffile{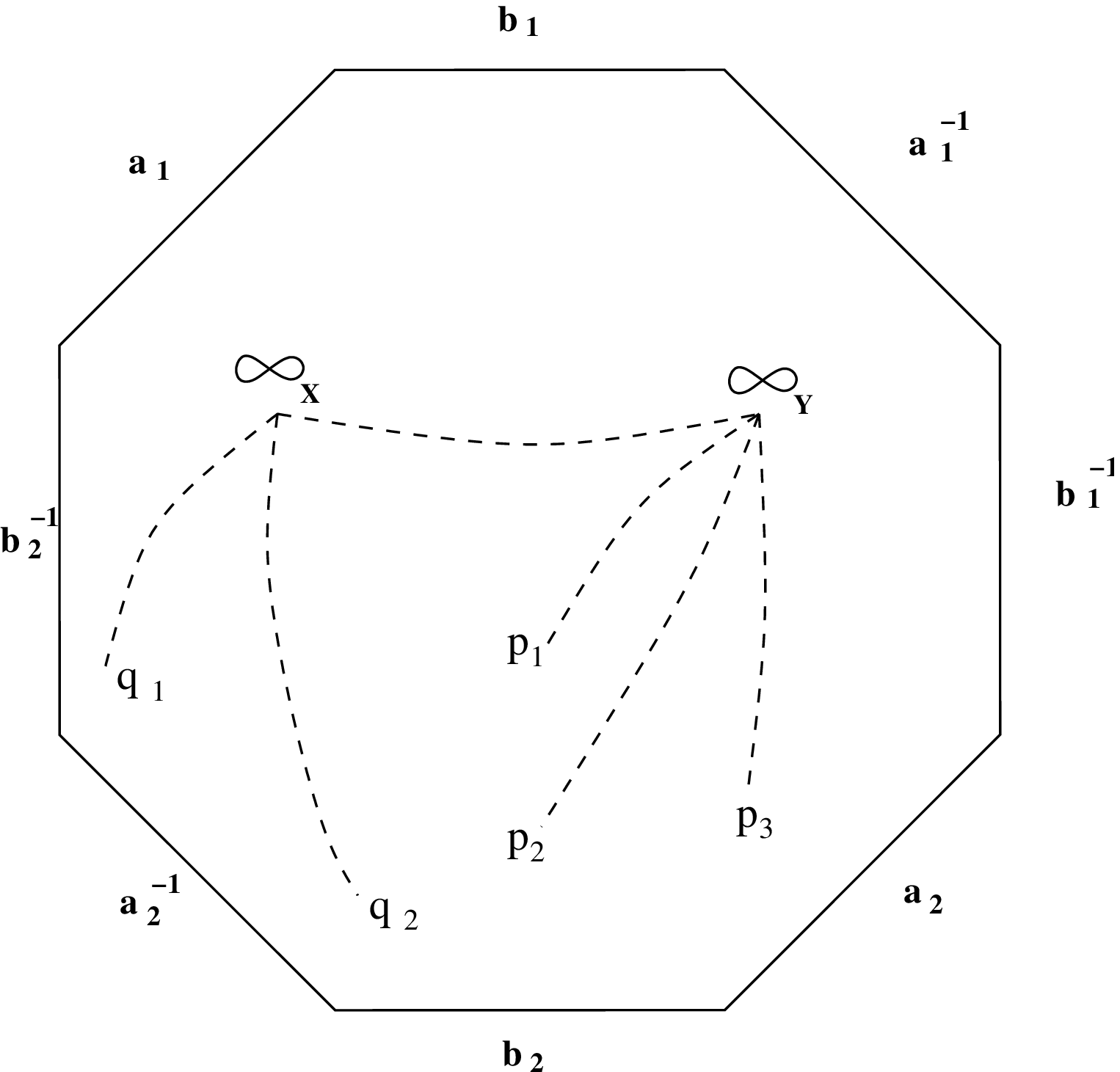}
\begin{center}
{\bf Figure 1} A visualization of an example of the dissection mentioned in the text
for a genus-$2$ curve.
\end{center}
}\\

In order to sketch how, let us first recall the {\bf thermodynamic
  identity}
\be
(\pa \Y)_\X{\rm d}\X = -(\pa \X)_\Y{\rm d}\Y\label{thermo}
\ee
where the subscript denotes the local coordinate to be kept fixed
under variation. As an example of the use of (\ref{thermo}) in
identifying the various differentials we consider a derivative
$\pa_{u_K}$.
From the defining relations for the coordinates (\ref{coord1}) we see
that
\be
(\pa_{u_{K,\infty}}\Y)_\X{\rm d}\X = \le\{\begin{array}{ll}
\ds \X^{K-1}{\rm d}\X + \mathcal O(\X^{-2}){\rm d}\X &\hbox{ near }\infty_\X\\
\ds \mathcal O(1){\rm d}\X & \hbox{ near } q_\alpha
\end{array}\ri.
\ee
has a  pole of order $K$ at $\infty_\X$ without residue. In order what
kind of singularity it has at $\infty_\Y$ we use (\ref{thermo})
followed by (\ref{coord1})
\bea
(\pa_{u_{K,\infty}}\Y)_\X{\rm d}\X = -(\pa_{u_K}\X)_\Y{\rm d}\Y = \le\{\begin{array}{ll}
 \mathcal O(\Y^{-2}){\rm d}\Y &\hbox{ near }\infty_\Y\\
\ds \mathcal O(1){\rm d}\Y & \hbox{ near } p_\alpha
\end{array}\ri.\ .
\eea
Therefore the differential $(\pa_{u_{K,\infty}}\Y)_\X{\rm d}\X$ has only a pole at
$\infty_\X$ and no residues: moreover it follows by differentiation of
(\ref{coord2}) that this differential is also normalized (i.e. with
vanishing $a$-cycles), which is sufficient to uniquely specify it.
It is then an exercise using the properties of $\Omega$ to see that
\be
(\pa_{u_{K,0}}\Y)_\X{\rm d}\X = -\res{\infty_\X} \frac {\X^{K}}{K}\Omega\ .
\ee
Following the same logic and similar reasoning  one can prove  the following formul\ae
\bea
\hbox{ First kind} && (\pa_{\epsilon_j}\Y)_\X{\rm d}\X = \omega_j= \frac
1{2i\pi}\oint_{b_j}\Omega\label{diff1}\\[10pt]
\hbox{Second kind} && \le\{\begin{array}{l}
\ds (\pa_{u_{K,\infty}}\Y)_\X{\rm d}\X = -\res{\infty_\X} \frac {\X^{K}}{K}\Omega=: \omega_{K,\infty}\\[10pt]
\ds (\pa_{u_{K,\alpha}}\Y)_\X{\rm d}\X = -\frac 1 K \res{q_\alpha}
(\X-Q_\alpha)^{-K}\Omega=: \omega_{K,\alpha} \\ [10pt]
\ds (\pa_{X_j}\Y)_\X{\rm d}\X =  \res{\xi_j}\Y \Omega=: \omega_{X_j} \\[10pt]
\ds (\pa_{Q_\alpha} \Y)_\X{\rm d}\X = \res{q_\alpha}
\le(V'_{1,\alpha}(\X) - \frac {u_{0,\alpha}}{(\X-Q_\alpha)} \ri)\Omega
\\[10pt]
\ds (\pa_{v_{J,\infty}}\Y)_\X{\rm d}\X = \res{\infty_\Y} \frac
    {\Y^{J}}{J}\Omega=:\omega_{\wt J,\infty}\\[10pt]
\ds (\pa_{v_{J,\alpha}}\Y)_\X{\rm d}\X = \frac 1 J \res{p_\alpha}
(\Y-P_\alpha)^{-J}\Omega = \omega_{\wt J,\alpha}\\[10pt]
  \ds (\pa_{Y_j}\Y)_\X{\rm d}\Y =  -\res{\eta_j} \X \Omega
    =:\omega_{Y_j}\\[10pt]
\ds (\pa_{P_\alpha} \Y)_\X{\rm d}\X =- \res{p_\alpha}
\le(V'_{2,\alpha}(\Y)-\frac {v_{0,\alpha}}{(\Y-P_\alpha)} \ri)\Omega
\end{array}\ri.\label{diff2}\\
\hbox{Third kind} &&
\le\{\begin{array}{l}
\ds( \pa_{u_{0,\alpha}} \Y)_\X{\rm d}\X = \int^{\infty_\X}_{q_\alpha}\Omega =:
\omega_{0,\alpha}\\
\ds (\pa_{v_{0,\alpha}} \Y)_\X{\rm d}\X = -\int^{\infty_\Y}_{p_\alpha}\Omega =:
\omega_{\wt 0,\alpha}\\
\ds (\pa_t \Y)_\X{\rm d}\X =
\int_{\infty_\Y}^{\infty_\X}\Omega=: \omega_0
\end{array}\ri . \label{diff3}
\eea
The only formul\ae\ above that need some further explanations are the
ones for the derivatives w.r.t. $X_j$ (or similarly $Y_j$);
from the asymptotic behavior (\ref{coord1}) in the local parameter
$z=\sqrt{\X-X_j}$ we have
\be
\pa_{X_j} \Y{\rm d}\X = \le[\frac {(\pa_{X_j} X_j)}2 \frac {\sqrt{-2R_j}}{z^3} + \frac
    {\pa\sqrt{-2R_j}}{z} + \mathcal O(1)\ri] 2z{\rm d}z  =
\frac {\sqrt{-2R_j}}{z^2}{\rm d}z + \mathcal O(1) = -{\rm d}\le(\frac
      {\sqrt{-2R_j}}{z} + \mathcal O(1)\ri) = \res{\xi_j} \Y\Omega
\ee
This proves that if $\pa=\pa_{X_j}$ then the differential has a double
pole at $\xi_j$ without residues: similar reasoning at the other
singularities and for the $a$-cycles of the differential force it to
be equal to the formula above in (\ref{diff2}).
\br
As explained in the introduction, we are also  interested to the restriction
of $\mathcal F$ to the subvariety of the moduli space defined by
\be
\pa_{\epsilon_j}\mathcal F
\le(V_1,V_2,t,\underline\epsilon(V_1,V_2,t)\ri)\equiv 0\ ,\ \ j=1,\dots,g.
\ee
Since on this subvariety the differentials $\X{\rm d}\Y, \Y{\rm d}\X$
have identically vanishing $b$-periods, the formulas for the
constrained derivatives that substitute (\ref{diff2},
\ref{diff3})\footnote{The equations (\ref{diff1}) do not make sense on
  the subvariety since $\epsilon$ are not independent coordinates any
  longer.} are the same with $\wt \Omega$ (\ref{dual}) replacing $\Omega$.
\er
Before writing the cross derivatives in a way which is
symmetric in $\X,\Y$,  we introduce some useful notation:  all the differentials (\ref{diff1}, \ref{diff2}, \ref{diff3}) are obtained by applying a suitable integral operator to one variable
of the Bergman kernel $\Omega$ according to the following table of translation
\bea
&& \begin{array}{lll|lll}
\ds\frac \pa{\pa u_{K,\infty}} &\mapsto& \ds\mathcal U_{K,\infty}:= - \frac 1{2i K \pi}
\oint_{\infty_\X}{\X^K}
&
\ds\frac \pa{\pa v_{J,\infty}} &\mapsto &\ds\mathcal V_{J,\infty}:=  \frac 1{2i J \pi}
\oint_{\infty_\Y}{\Y^J}
\cr
\ds\frac \pa{\pa u_{0,\alpha}} &\mapsto&\ds \mathcal U_{0,\alpha}:=
\slint_{q_\alpha}^{\infty_\X}
&
\ds\frac \pa{\pa v_{0,\alpha}} &\mapsto& \ds\mathcal V_{0,\alpha}:=
-\slint_{p_\alpha}^{\infty_\Y}
\cr
\ds\frac \pa{\pa u_{K,\alpha}} &\mapsto& \ds\mathcal U_{K,\alpha}:= - \frac
1{2i K \pi}
\oint_{q_\alpha}\frac 1{(\X-Q_\alpha)^K}
&
\ds\frac \pa{\pa v_{J,\alpha}} &\mapsto&\ds \mathcal V_{J,\alpha}:=   \frac
1{2i J \pi}
\oint_{p_\alpha}\frac 1{(\Y-P_\alpha)^J}
\cr
\ds\frac \pa{\pa X_j} &\mapsto & \ds\mathcal R_J:=   \frac 1
      {2i\pi}\oint_{\xi_j} \Y
&
\ds\frac \pa{\pa Y_j} &\mapsto &\ds \mathcal S_J:= - \frac 1
      {2i\pi}\oint_{\eta_j} \X
%
\end{array}\cr
&&\begin{array}{lll}
\ds \frac \pa{\pa t} &\mapsto& \ds\mathcal T:= \slint_{\infty_\Y}^{\infty_\X}\cr
\ds\frac \pa{\pa \epsilon_j} &\mapsto&\ds \mathcal E_j:= \frac
1{2i\pi}\oint_{b_j}\ .
\end{array}
\label{table}
\eea
All the differentials (\ref{diff1}, \ref{diff2},
\ref{diff3}) are obtained by applying the corresponding integral
operator in (\ref{table}) to the Bergman bidifferential $\Omega$.\\
In order to write the cross derivatives let us choose two coordinates
and denote by $\pa_1,\pa_2$ the corresponding
derivatives and by $\int_{\pa_1},\int_{\pa_2}$ the corresponding
integral operator as per the table (\ref{table}): then we have
\be
\pa_1\pa_2\mathcal F = \pa_1 \int_{\pa_2} \Y{\rm d}\X = \int_{\pa_2}
(\pa_1 \Y)_\X{\rm d}\X = \int_{\pa_2}\int_{\pa_1} \Omega\ .
\ee
The important and conclusive remark now is that the order of the
action of the integral operators appearing in the list (\ref{table}) on $\Omega$ is immaterial because the
kernel $\Omega$ is symmetric and -more importantly- because its
residue on the diagonal is zero. This means that in exchanging two
integral operators one may in fact acquire the integral of a total
differential which is going to cancel either by integration or against
the regularization. To illustrate the point we make two examples.
\bx
Consider two coordinates $u_{K,\alpha},v_{J,\beta}$: then
\be
\pa_{u_{K,\alpha}}\pa_{v_{J,\beta}} \mathcal F = \mathcal V_{J,\beta}
\mathcal U_{K,\alpha}\Omega\ .
\ee
In this case the two integral operators involve either residues (for
$K>0$) or (regularized) integrals. Either way the contours do not
intersect and the double integral is independent of the order.
\ex
\bx
Consider the derivatives $\pa_{u_{0,\alpha}}$ and
$\pa_{u_{K,\alpha}}$; in this case the integral operators do involve
intersecting contours, hence care must be exercised
\be
\pa_{u_{0,\alpha}}\pa_{u_{K,\alpha}}\mathcal F = \frac
1{2iK\pi}\oint_{q_\alpha}
(\X-Q_\alpha)^{-K}(\zeta)\slint_{\infty_\X}^{q_\alpha}
\Omega(\zeta,\xi)\ .
\ee
The inner integral in fact does not need any regularization, so we
have
\bea
\pa_{u_{0,\alpha}}\pa_{u_{K,\alpha}}\mathcal F = \frac
1{2iK\pi}\oint_{q_\alpha}
(\X-Q_\alpha)^{-K}(\zeta)\int_{\infty_\X}^{q_\alpha}
\Omega(\zeta,\xi) =
\\
= \lim_{\epsilon\to q_\alpha}  \frac
1{2iK\pi}\int_{\infty_\X}^{\epsilon} \oint_{q_\alpha}
(\X-Q_\alpha)^{-K}(\zeta)\Omega(\zeta,\xi) -
\frac 1 K(\X(\epsilon)-Q_\alpha)^{-K} =
\\
= \frac
1{2iK\pi}\slint_{\infty_\X}^{q_\alpha}\oint_{q_\alpha}
(\X-Q_\alpha)^{-K}(\zeta)\Omega(\zeta,\xi)=
 \pa_{u_{0,\alpha}}\pa_{u_{K,\alpha}}\mathcal F \ .
\eea
(The exchange of the order of the integrals gives a $-2i\pi\delta$ supported at the
intersection of the contours of integration).
\ex
\bt
\label{main}
The free energy is given by the formula (we set
$u_{K,0}:=u_{K,\infty},\ v_{J,0}:=v_{J,\infty},\ u_{0,0} =
u_{0,\infty}:= v_{0,0} = v_{0,\infty}:=0$
for uniformity in the formul\ae)
\be
2 \mathcal F = \sum_{\alpha=0} \sum_{K=0}^{d_{1,\alpha}} u_{K,\alpha}
U_{K,\alpha} +   \sum_{\alpha=0} \sum_{J=0}^{d_{2,\alpha}} v_{J,\alpha}
V_{J,\alpha} +t\mu  +\sum_{j=1}^g \epsilon_j \Gamma_j+
\le\{
\begin{array}{l}
\ds  \frac 1 2 \sum_{\zeta\in \mathcal D_\X} \res{\zeta}
\Y^2 \X{\rm d}\X\cr
\ds  \frac 1 2 \sum_{\zeta\in \mathcal D_\Y} \res{\zeta}
\X^2 \Y{\rm d}\Y\cr
\end{array}\ri.
 \ ,\label{freeenergy}
\ee
where
\bea
\mathcal D_\X:= \{\infty_\X,q_\alpha,\xi_j,\ \alpha=1,\dots;\
j=1,\dots\} \\
 \mathcal D_\Y:= \{\infty_\Y,p_\alpha,\eta_j,\ \alpha=1,\dots;\
 j=1,\dots \}
\eea
(see definitions of the properties of the points appearing here
at the beginning of Sect. \ref{setting})\footnote{ The set $\mathcal
  D_\X$ is the support of the pole-divisor of $\Y$ less the point
  $\infty_\Y$, and viceversa for $\mathcal D_\Y$.}
\et
{\bf Proof}.
First of all note that the expression is symmetric in the r\^oles of
$\X,\Y$  after integration  by parts and moving  the residues to the other poles
\be
 \frac 1 2 \sum_{\zeta\in \mathcal D_\X} \res{\zeta} \Y^2 \X{\rm d}\X=
 - \frac 1 2 \sum_{\zeta\in \mathcal D_\Y} \res{\zeta} \Y^2 \X{\rm d}\X=  \frac
 1 2 \sum_{\zeta\in\mathcal D_\Y} \res{\zeta}\X^2\Y{\rm d}\Y\ ,
\ee
where we have used that $\mathcal D_\X\cup \mathcal D_\Y$ is the set
of all poles of the differential $\Y^2\X{\rm d}\X$.
Now, the proposed expression is nothing but
\bea
2\mathcal F &=&\sum_{\alpha=0} \sum_{K=0}^{d_{1,\alpha}} u_{K,\alpha}
\mathcal U_{K,\alpha}(\Y{\rm d}\X) -  \sum_{\alpha=0} \sum_{J=0}^{d_{2,\alpha}} v_{J,\alpha}
\mathcal V_{J,\alpha} (\X{\rm d}\Y) +  t\mathcal T (\Y{\rm d}\X)  +\\
&&+\sum_{j=1}^g \epsilon_j \mathcal E_j( \Y{\rm d}\X ) +  \frac 1 2
\sum_{\zeta\in \mathcal D_\X} \res{\zeta} \Y^2 \X{\rm d}\X  - t\sum v_{0,\alpha}
\eea
Suppose we compute a derivative w.r.t. $u_{R,\beta}$: using the list
of differentials (\ref{diff1}, \ref{diff2}, \ref{diff3}) and moving
the computation of residues over to $\mathcal D_\Y$ --for convenience--
before the differentiation, we have
\bea
2\pa_{u_{R,\beta}}\mathcal F& =& \overbrace{\mathcal
  U_{R,\beta}(\Y{\rm d}\X)}^{=U_{R,\beta}}
 + \sum_{\alpha=0} \sum_{K=0}^{d_{1,\alpha}} u_{K,\alpha}
\mathcal U_{K,\alpha}\le(\mathcal U_{R,\beta}\Omega \ri) + \\
&& -   \sum_{\alpha=0} \sum_{J=0}^{d_{2,\alpha}} v_{J,\alpha}
\mathcal V_{J,\alpha}\le(-\mathcal U_{R,\beta}\Omega \ri) + t\mathcal
  T\le(\mathcal U_{R,\beta}\Omega \ri)    + \sum_{j=1}^g \epsilon_j
  \mathcal E_j\le(\mathcal U_{R,\beta}\Omega \ri) -
  \sum_{\zeta\in \mathcal D_\Y} \res{\zeta} \Y \X \mathcal U_{R,\beta}(\Omega) =\\
&&\hspace{-2cm}= U_{R,\beta} + \mathcal U_{R,\beta}\le( \sum_{\alpha=0} \sum_{K=0}^{d_{1,\alpha}} u_{K,\alpha}
\mathcal U_{K,\alpha}\le(\Omega \ri) +   \sum_{\alpha=0} \sum_{J=0}^{d_{2,\alpha}} v_{J,\alpha}
\mathcal V_{J,\alpha} \le(\Omega \ri) + t\mathcal
  T\le(\Omega \ri)    + \sum_{j=1}^g \epsilon_j
  \mathcal E_j\le(\Omega \ri)  - \sum_{\zeta\in \mathcal D_\Y}
  \res{\zeta} \X\Y\Omega\ri) \ .
\eea
Note that the operator $\mathcal U_{R,\beta}$ involves residues at one
of the points of $\mathcal D_\X$ and hence commutes with the other
residues when acting on the (singular) kernel $\Omega$ also for the
last term involving residues at $\mathcal D_\Y$.\par
From the properties of $\Omega$ and the definitions of the integral
operators it follows that the differential acted upon by $\mathcal
U_{R,\beta}$ is precisely $\Y{\rm d}\X$, namely
\be
 \Y{\rm d}\X =\sum_{\alpha=0} \sum_{K=0}^{d_{1,\alpha}} u_{K,\alpha}
\mathcal U_{K,\alpha}\le(\Omega \ri) +   \sum_{\alpha=0} \sum_{J=0}^{d_{2,\alpha}} v_{J,\alpha}
\mathcal V_{J,\alpha} \le(\Omega \ri) + t\mathcal
  T\le(\Omega \ri)    + \sum_{j=1}^g \epsilon_j
  \mathcal E_j\le(\Omega \ri) 
-\sum_{\zeta\in \mathcal D_\Y} \res{\zeta} \X\Y\Omega \ .
\ee
This can be seen by analyzing the singular behavior near the poles and
the $a$-periods of both sides of the equality and verifying that they
are the same\footnote{ One should use that the behaviour near a pole of the last
  term in the LHS is, e.g.
\bes
\sum_{\zeta\in \mathcal D_\Y}
  \res{\zeta} \X\Y\Omega \mathop{\sim}_{p_\alpha} - {\rm d}\le(\Y
  V_{2,\alpha}'(\Y)\ri)\ .
\ees}
whence we have the desired conclusion of this part of the proof. The other
derivatives are treated in completely parallel way. \par
The derivatives w.r.t. $X_j,Y_j$ are a little different because there
is no explicit dependence of $\mathcal F$ from  these
coordinates. However this produces the correct result since, for example
\bea
2\pa_{X_\ell}\mathcal F& =& \sum_{\alpha=0} \sum_{K=0}^{d_{1,\alpha}} u_{K,\alpha}
\mathcal U_{K,\alpha}\le(\mathcal R_\ell\Omega \ri) -   \sum_{\alpha=0} \sum_{J=0}^{d_{2,\alpha}} v_{J,\alpha}
\mathcal V_{J,\alpha}\le(-\mathcal R_\ell\Omega \ri) + t\mathcal
  T\le(\mathcal R_\ell\Omega \ri)    +
\\
&&
 + \sum_{j=1}^g \epsilon_j
  \mathcal E_j\le(\mathcal R_\ell\Omega \ri) 
-\sum_{\zeta\in \mathcal D_\Y} \res{\zeta} \X\Y \mathcal R_\ell (\Omega )=\\
&=& \mathcal R_l (\Y{\rm d}\X)\ ,
\eea
which is consistent with our definitions (\ref{firstders}).\par
As a final case we compute the derivative w.r.t. $Q_\alpha$: here
some care should be paid to the commutation of the derivative with the
integral operators. Indeed $\pa_{Q_{\alpha}}$ does not commute with
the integral operators $\mathcal U_{K,\alpha}, K=0,\dots$ but instead
we have
\bea
[\pa_{Q_{\alpha}}, \mathcal U_{K,\alpha}] &=& K\mathcal
U_{K+1,\alpha},\ K=1,\dots\label {commutat1}\\[1pt]
[\pa_{Q_{\alpha}}, \mathcal U_{0,\alpha}] &=&  \mathcal U_{1,\alpha} \label {commutat2}\ .
\eea
While (\ref{commutat1}) is rather obvious from the definition of the
integral operator, some explanation is necessary for
(\ref{commutat2}). Expanding $\slint^{\infty_\X}_\epsilon \Y{\rm d}\X$ in the local parameter at
$q_\alpha$ we have
\bea
\slint^{\infty_\X}_\epsilon \Y{\rm d}\X = -V_{1,\alpha}(\X(\epsilon)) + c_0 + c_1 z_\alpha + \mathcal
O(z_{\alpha}^2) + V_{1,\alpha}(\X(\epsilon))
\eea
Therefore we have
\bea
\pa_{Q_{\alpha}} \slint^{\infty_\X}_{q_\alpha} \Y{\rm d}\X = \pa_{Q_\alpha}\le(
\lim_{\epsilon\to q_\alpha} \int_\epsilon \Y{\rm d}\X +
V_{1,\alpha}(\X(\epsilon)) \ri) = \pa_{Q_{\alpha}} c_0
\eea
Viceversa (recalling that $\pa_{Q_\alpha} z_\alpha = -1$)
\bea
 \slint^{\infty_\X}_{q_\alpha}(\pa_{Q_\alpha} \Y)_\X{\rm d}\X& =&
 \lim_{\epsilon\to q_\alpha} \int^{\infty_\X}_\epsilon (\pa_{Q_\alpha}
 \Y)_\X{\rm d}\X  =\cr
&=&  \lim_{\epsilon\to q_\alpha} \le(\pa_{Q_\alpha} c_0 - c_1 +
 \pa_{Q_{\alpha}} c_1 z_\alpha + \mathcal O(z_\alpha^2)\ri) = \pa_{Q_\alpha} c_0 - c_1
\eea
This shows that
\be
\le[\pa_{Q_\alpha},\slint_{q_\alpha}\ri] = -\frac 1{2i\pi}
\oint_{q_\alpha} \frac 1{\X-Q_\alpha} = \mathcal U_{1,\alpha}\ .
\ee
Using this and computing the derivative of $\mathcal F$ we obtain the
desired result
\be
\pa_{Q_{\alpha}}\mathcal F = \res{q_\alpha} V'_{1,\alpha}(\X)\Y{\rm d}\X\ .
\ee
Finally, while the reasoning is mostly similar, the $t$ derivative has
an additional technical difficulty. First of all we have
\bea
\pa_t\slint_{\infty_\Y}^{\infty_\X} \Y{\rm d}\X =
\slint_{\infty_\Y}^{\infty_\X} \slint_{\infty_\Y}^{\infty_\X} \Omega +
1\ .
\eea
The reason of the additional $+1$ is the fact that the local
parameters near the two poles are different functions (here we set for
brevity $t_\X=t+\sum u_{0,\alpha},\ t_\Y = t + \sum v_{0,\alpha}$) 
\bea
\pa_t\slint_{\infty_\Y}^{\infty_\X} \Y{\rm d}\X
=\lim_{\epsilon\to\infty_\Y\atop \rho\to\infty_\X}\pa_t \le[
\int_\epsilon^\rho\Y{\rm d}\X -\le( V_{1,\infty}(\X)-t_\X
\ln (\X)\ri)_{\rho} + \le(\Y V_{2,\infty}'(\Y)-V_{2,\infty}(\Y) -
t_\Y\ln(\Y)\ri)_\epsilon
\ri]=\cr
= \lim_{\epsilon\to\infty_\Y\atop \rho\to\infty_\X}\bigg[ 
\int_\epsilon^\rho\int_{\infty_\Y}^{\infty_\X}\Omega +\ln(\X(\rho))
-\ln(\Y(\epsilon)) + \le(\Y V_{2,\infty}''(\Y)-\frac {t_\Y}
\Y\ri)
\mathop{ (\pa_t \Y)_\X}^{\frac {{\rm d}\Y}{\Y{\rm d}\X} + \dots\atop ||}
\bigg] =\cr
= \slint_{\infty_\Y}^{\infty_\X} \slint_{\infty_\Y}^{\infty_\X} \Omega
+ 1\ .
\eea
 Moreover, whether we sum at the
poles in $\mathcal D_\X$ or $\mathcal D_\Y$, we need to interchange
the order of the following residue/integral
\bea
\res{\infty_\Y} \X\Y \int_{\infty_\Y}^{\infty_\X}\!\!\!\!\!\Omega =
\lim_{\epsilon\to \infty_\Y} \int^{\infty_\X}_\epsilon
\res{\infty_\Y}\X\Y\Omega -
\Y(\epsilon)\hspace{-1.45cm}\mathop{\X(\epsilon)}^{V'_{2,\infty}(\Y) - (t+\sum
  v_{0,\alpha})\Y^{-1} + \dots\atop ||}
\hspace{-1.4cm} =
\slint_{\infty_\Y}^{\infty_\X}\res{\infty_\Y}\X\Y\Omega  + t +
\sum_{\alpha} v_{0,\alpha}\ .
\eea
Putting it all together we find
\be
2\pa_t\mathcal F  = (\mathcal T(\Y{\rm d}\X)+t -\sum_{\alpha}v_{0,\alpha})+ \mathcal T (\Y{\rm d}\X)  - t -\sum_\alpha
v_{0,\alpha} = 2\mu.
\ee
The other derivatives w.r.t. to the moduli $v_{J,\alpha}, Y_j,
P_\alpha$  are computed in similar way by first rewriting the
expression for $\mathcal F$  equivalently in the symmetric way
w.r.t. the exchange of r\^oles of $\X,\Y$.
Q.E.D.\par\vskip 5pt
\bc
The Free energy satisfies the following scaling constraints 
\bea
2\mathcal F &=& \mathbb V_\Y \mathcal F + \sum_{1\leq \alpha< \beta} v_{0,\alpha}v_{0,\beta} + t\sum_{\alpha\geq 1
}v_{0,\alpha} + \frac {t^2}2  \\
2\mathcal F &=& \mathbb V_\X\mathcal F +
\sum_{1\leq \alpha< \beta} u_{0,\alpha}u_{0,\beta} + t\sum_{\alpha\geq 1
}u_{0,\alpha} + \frac {t^2}2 
\eea
where 
\bea
\mathbb V_\Y&:=& \sum_{\alpha\geq 0} \sum_{K\geq 0} u_{K,\alpha} \frac
\pa{\pa u_{K,\alpha}}  +\sum_{J=1}^{d_{2,\infty}} (1-J)v_{J,\infty}
\frac \pa{\pa v_{J,\infty}} + \sum_{\alpha\geq
  1}\le( P_{\alpha} \frac \pa{\pa P_\alpha} 
+ \sum_{J=0}^{d_{2,\alpha}} (J+1)v_{J,\alpha}\frac \pa{\pa
  v_{J,\alpha}} \ri)+
\cr
&&  + \sum_j Y_j\frac \pa {\pa Y_j} +t\frac \pa{\pa t} + \sum_{j=1}^g
  \epsilon_j \frac \pa {\pa \epsilon_j}
\cr
\mathbb V_\X&:=&  \sum_{\alpha\geq 0} \sum_{J\geq 0} v_{J,\alpha} \frac
\pa{\pa v_{J,\alpha}}  +\sum_{K=1}^{d_{1,\infty}} (1-K)u_{K,\infty}
\frac \pa{\pa u_{K,\infty}} + \sum_{\alpha\geq
  1}\le( Q_{\alpha} \frac \pa{\pa Q_\alpha} 
+ \sum_{K=0}^{d_{1,\alpha}} (K+1)u_{K,\alpha}\frac \pa{\pa
  u_{K,\alpha}} \ri)+
\cr
&&   + \sum_j X_j\frac \pa {\pa X_j} +t\frac \pa{\pa t} + \sum_{j=1}^g
  \epsilon_j \frac \pa {\pa \epsilon_j}
\eea
Note that these formul\ae\ give other representations of the free energy
in terms of its first derivatives defined independently in
(\ref{firstders}).
Moreover any convex linear combination will give another representation.
\ec
{\bf Proof.} 
The formul\ae\ can be obtained by explicitly computing the residues of
$\Y^2\X{\rm d}\X$ at the various points or by the following
straightforward argument. 
Consider the new functions $\wt\X:=\X$ and $\wt \Y = {\rm e}^c \Y$:
the new free energy $\wt{\mathcal F}$ will be given by the same
formula  (\ref{freeenergy}) in terms of the new objects. Taking
$\frac {{\rm d}}{{\rm d}c}\bigg|_{c=0}$ gives the first formula.
Some particular care has to be paid to the regularizations which
involve subtraction of logarithms.\\
The second formula  is obtained in a symmetric way.
Q.E.D.\par\vskip 6pt

If we denote  by $\int_\pa$ the integral operator
associated to a derivative $\pa$, the formulas for the second order
derivatives are written concisely
\be
\pa_1\pa_2 \mathcal F = \int_{\pa_1}\int_{\pa_2}\Omega +
\delta_{\pa_1,t}\delta_{\pa_1,\pa_2}
\ee
In other words\footnote{We could dispose of the last term (enters only in $\pa^2_t\mathcal F$)
by subtracting $\frac 1 2 t^2$; this would change the $t$-derivative
$\mu\to\mu+t$ making the formula for the first derivatives slightly
different. Note that this does not affect the derivatives of order $3$
and higher.}
 the Bergman kernel is the universal kernel for
computing the second derivatives of the free energy  and hence  the two-point correlation
functions of the matrix model in the planar limit.\par
The third order correlation functions were computed in \cite{F2}
for the case of polynomial potentials: since the reasoning is
identical we only report the result.
The key ingredient there is the formula that allows you to find the
variation of the Bergman kernel under infinitesimal change of the
deformation parameters. The formulas can be summarized as follows
\bea
(\pa \Omega)_\X(\xi,\eta) &=&-\int_{\rho,\pa} \sum_k \res{\zeta=x_k} \frac
    {\Omega(\xi,\zeta)\Omega(\rho,\zeta)\Omega(\eta,\zeta)}{{\rm
    d}\Y(\zeta){\rm d}\X(\zeta)} = -\sum_k \res{\zeta=x_k} \frac
    {\Omega(\xi,\zeta)\omega_\pa(\zeta)\Omega(\eta,\zeta)}{{\rm
    d}\Y(\zeta){\rm d}\X(\zeta)} \cr
(\pa \Omega)_\Y(\xi,\eta) &=&\int_{\rho,\pa} \sum_k \res{\zeta=y_k} \frac
    {\Omega(\xi,\zeta)\Omega(\rho,\zeta)\Omega(\eta,\zeta)}{{\rm
    d}\Y(\zeta){\rm d}\X(\zeta)}=\sum_k \res{\zeta=y_k} \frac
    {\Omega(\xi,\zeta)\omega_\pa(\zeta)\Omega(\eta,\zeta)}{{\rm
    d}\Y(\zeta){\rm d}\X(\zeta)}\label{rauchforms}
\eea
where $x_k$ and $y_k$ denote -respectively- {\em all} the critical points of $\X$
and $\Y$ other than $\infty_\Y,\infty_\X$ (namely ${\rm d}\X(x_k)=0$,
${\rm d}\Y(y_j)=0$).
These formulas follow from Rauch variational formula
\cite{rauch,korotkin}.
Note that ${\rm d}\Y{\rm d}\X$ in the denominator has simple poles at
the $\xi_j,\eta_j$, hence the residues at these points  do not contribute to the sum except
for the cases where $\omega_\pa$ has a (double) pole at one of those
points, namely only for the cases $\pa = \pa_{X_j},\ \pa_{Y_j}$.
\par
The final formul\ae\ for the third derivatives are simpler if we
introduce the two kernels
\bea
\Omega^{(3)}_\X(\zeta_1,\zeta_2,\zeta_3)&:=& -\sum_k \res{\zeta=x_k} \frac
    {\Omega(\zeta_1,\zeta)\Omega(\zeta_2,\zeta)\Omega(\zeta_3,\zeta)}{{\rm
    d}\Y(\zeta){\rm d}\X(\zeta)}\\
\Omega^{(3)}_\Y(\zeta_1,\zeta_2,\zeta_3)&:=& \sum_k \res{\zeta=y_k} \frac
    {\Omega(\zeta_1,\zeta)\Omega(\zeta_2,\zeta)\Omega(\zeta_3,\zeta)}{{\rm
    d}\Y(\zeta){\rm d}\X(\zeta)}
\eea
This way one obtains
\bea
\pa_{u_{K,\alpha}}\pa_{u_{J,\beta} }\pa \mathcal F = \int_\pa \mathcal U_{K,\alpha}\mathcal U_{J,\beta}
\Omega^{(3)}_\X\ ,\qquad \pa_{v_{K,\alpha}}\pa_{v_{J,\beta}}\pa
\mathcal F = \int_\pa \mathcal V_{K,\alpha}\mathcal V_{J,\beta}
\Omega^{(3)}_\Y\label{third1}\\
\pa_{u_{K,\alpha}}\pa_{t}\pa_t \mathcal F = \mathcal U_{K,\alpha}\mathcal T\mathcal
T\Omega^{(3)}_\X\ ;\qquad \pa_{v_{J,\alpha}}\pa_{t}\pa_t \mathcal F = \mathcal V_{J,\alpha}\mathcal T\mathcal
T\Omega^{(3)}_\Y.\label{third2}
\eea
For all other third order derivatives one can use either kernels:
\bea
\pa_1\pa_2\pa_3\mathcal F =
\int_{\pa_1}\int_{\pa_2}\int_{\pa_3}\Omega^{(3)}_\Y =
\int_{\pa_1}\int_{\pa_2}\int_{\pa_3}\Omega^{(3)}_\X\label{third3}
\eea
It should be clear to the reader that these formul\ae\ translate to
{\em residue formulas} in the spirit of \cite{Kric, dubrovin}. For
example
\be
\pa_{\epsilon_j}\pa_{\epsilon_k}\pa_{\epsilon_\ell} \mathcal F =  -\sum_k \res{\zeta=x_k} \frac
    {\omega_j\omega_k\omega_\ell}{{\rm d}\Y{\rm d}\X} =  \sum_k \res{\zeta=y_k} \frac
    {\omega_j\omega_k\omega_\ell}{{\rm d}\Y{\rm d}\X}
\ee
We should stress, however, that  although the present moduli space can
be embedded as a submanifold of the moduli space considered in
\cite{Kric}, the coordinates $(u_k,v_J)$ that are relevant to the matrix-model
applications are of a different sort and -resultingly- the free energy it is not the
same function as the tau function of the Whitham hierarchy in \cite{Kric}.
\section{Residue formulas for  higher derivatives:
  extended Rauch-variational formul\ae}
\label{second}
It is clear from the previous review of the material that in order to
compute any further variation we must be able to find the variation of
the kernels $\Omega^{(3)}_{\Y}$ and $\Omega^{(3)}_\X$: this step will
produce {\em three} kernels
\be
\Omega^{(4)}_{\Y\Y}\ ,\ \Omega^{(4)}_{\Y\X}\ ,\
\Omega^{(4)}_{\X\X}\ ,
\ee
according to which variable $\Y$ or $\X$ we keep fixed under the new
variation. The reason of this plethora is essentially that the
variations of the basic differentials $\int_\pa \Omega$ are performed
more easily either at $\Y$ or $\X$ fixed: for instance if we
compute the variation of $\omega_K= \mathcal U_K(\Omega)$ at $\X$-fixed
we obtain
\be
(\pa\omega_{K,\infty})_\X (\xi)= \res{\infty_\X}\frac {\X^K}K (\pa\Omega)_\X =
-\sum_{k}\res{\zeta=x_k} \frac {\omega_K(\zeta)
  \omega_\pa(\zeta)\Omega(\xi,\zeta)}{{\rm d}\Y(\zeta){\rm d}\X(\zeta)}
\label{acca}
\ee
whereas
\be
(\pa\omega_{K,\infty})_\Y = \res{\infty_\X} \le(\X^{K-1}(\pa \X)_\Y \Omega +\frac {\X^K}K (\pa\Omega)_\Y \ri)\label{cacca}
\ee
This is in fact a manifestation of the thermodynamic identity for
differentials
\bl
\label{lemma2}
The variation of a differential at $\Y$ and $\X$ fixed are related by
the following formula
\be
(\pa \omega)_\Y = (\pa\omega)_\X + {\rm d}\le(\frac \omega{{\rm d}\X} (\pa
  \X)_\Y\ri) =(\pa\omega)_\X - {\rm d}\le(\frac{ \omega\omega_\pa}{{\rm d}\X{\rm d}\Y}\ri)
\ee
\el
{\bf Proof.}
Writing $\omega = f {\rm d}\Y = g {\rm d}\X$ we have
\be
(\pa\omega)_\Y = \le[(\pa g)_\X + \frac {{\rm d}g}{{\rm d}\X}(\pa
  \X)_\Y\ri] {\rm d}\X+ g{\rm d}(\pa \X)_\Y = (\pa\omega)_\X + {\rm d}
\le(g(\pa \X)_\Y\ri)\ .
\ee
Since $g = \omega/{\rm d}\X$ and $(\pa \X)_\Y = -\omega_\pa/{\rm d}\Y$ we
have the assertion.
Q.E.D.\par\vskip 5pt
Using Lemma \ref{lemma2} and trading the residues at the $x_k$ over to
the others (at the $y_\ell,\xi,\infty_\X$) one can check directly that
the formul\ae\ (\ref{acca}, \ref{cacca}) are consistent.\par
It should also be clear that the variation of the numerators of
$\Omega^{(3)}_{\Y,\X}$ are obtained by simply applying the product rule
and the previously listed appropriate Rauch formul\ae. The only new
ingredient is the variation of the denominator of
$\Omega^{(3)}_{\Y,\X}$ as explained below. \par
Suppose we want to perform a variation $\pa$  at $\X$ fixed of one of the two
kernels; when we need to compute the variation of the denominator we
need a formula for $\pa \frac 1 {{\rm d}\Y}$.  We should think of the
expression $\frac 1{{\rm d}\Y}$ as a meromorphic vector field on the
  Riemann surface and the variation is the vector field
\be
\pa\le(\frac 1 {{\rm d}\Y}\ri)_\X = -\frac {{\rm d}((\pa \Y)_\X)}{{\rm
    d}\Y^2} = - \frac 1{{\rm d}\Y^2} {\rm d}\le(\frac{\omega_\pa}{{\rm
    d}\X}\ri)
\label{vectf}
\ee
Now, the differential of the function $\frac{\omega_\pa}{{\rm d}\X}$
can be expressed as a residue using -once more- the Bergman kernel
according to the following
\bl
\label{lemma1}
Let $F$ be a (local) meromorphic function: then the differential ${\rm
  d}F$ can be obtained by
\be
{\rm d}F(\xi) = \res{\zeta=\xi} \Omega(\zeta,\xi)F(\zeta)\ .
\ee
\el
The proof is very simple using a local parameter near the point $\xi$
and the asymptotic expansion of the Bergman kernel.\par
Combining Lemma \ref{lemma1} with (\ref{vectf}) we have the new
  variational formula
\bea
\pa\le(\frac 1 {{\rm d}\Y}\ri)_\X\bigg|_{\xi} & =& -\frac 1{{\rm
  d}\Y^2(\xi)}\res{\zeta=\xi}\Omega(\zeta,\xi)\frac{\omega_\pa(\zeta)}{{\rm
    d}\X(\zeta)} \cr
\pa\le(\frac 1 {{\rm d}\X}\ri)_\Y\bigg|_{\xi}  &= &\frac 1{{\rm
  d}\X^2(\xi)}\res{\zeta=\xi}\Omega(\zeta,\xi)\frac{\omega_\pa(\zeta)}{{\rm
    d}\Y(\zeta)}
 \label{vectfvar}
\eea
where the different sign in the second formula is due to the fact that
$(\pa \X)_\Y = -\omega_\pa /{\rm d}\Y$.\\
Let us summarize the {\bf rules of the calculus}:
\begin{enumerate}
\item The variations of any differential can be performed at $\X$ or
  $\Y$ fixed, the two being related by Lemma \ref{lemma2}.
\item The variations at $\X$-fixed of the vector fields $1/{\rm d}\Y$
  and viceversa  are given by eq. (\ref{vectfvar}).
\item The variations of the Bergman bidifferential $\Omega$ are given
  by eqs. (\ref{rauchforms})
\end{enumerate}
The choice of variable to be kept fixed $\Y$ vs. $\X$ is ultimately
immaterial. However formul\ae\ can take on a significantly more
involved form if one chooses the ``wrong'' way of differentiation. We
are going to practice this calculus and compute the fourth order
derivatives explicitly. This will also provide us with relevant
formul\ae\  for the four point correlators of the planar limit of the
two-matrix model.
\subsection{Fourth order}
To illustrate the method we compute the fourth derivatives
w.r.t. $u_{K,\alpha},u_{L,\beta},u_{M,\gamma},u_{N,\delta}$ (which we
will denote in short hand by subscripts $_M,_N,_L,_K$ only). We start from the expression for the third derivative
\be
\pa_M\pa_N\pa_L\mathcal F := \mathcal F_{M,N,L}  = -\sum \res{\xi=x_k}
\frac{\omega_M\omega_L\omega_N}{{\rm d}\Y{\rm d}\X}\ .
\ee
It is quite obvious from the considerations around  Eq. (\ref{acca}) that
the extra derivative is most easily computed at $\X$-fixed: 
\bea
\pa_K {\mathcal F}_{M,N,L} =-
 \sum_{k} \res{\xi=x_k}
\frac{(\pa_K\omega_M)_\X\omega_L\omega_N}
{{\rm d}\Y{\rm d}\X}
 - \  (M\leftrightarrow L)-(M\leftrightarrow N)\    
+ \sum \res{\xi=x_k} \frac
{\omega_L\omega_M\omega_N}{{\rm d}\Y{\rm d}\X} \frac {{\rm d}(\pa_K
  \Y)_\X}{{\rm d}\Y}
\eea
Using now Lemma \ref{lemma1} and the variational formul\ae\
(\ref{rauchforms}) we obtain
\bea
\pa_K {\mathcal F}_{M,N,L}
=
\sum_{k} \res{\xi=x_k}
\frac{\omega_L(\xi)\omega_N(\xi)}
{{\rm d}\Y(\xi){\rm d}\X(\xi)}\le( \sum_{\ell} \res{\zeta=x_\ell}
\frac
   {\omega_M(\zeta)\omega_{K}(\zeta)\Omega(\xi,\zeta)}{{\rm d}\Y(\zeta){\rm
       d}\X(\zeta)}\ri)+ \   (M\leftrightarrow L)+(M\leftrightarrow N ) +\\
+ \sum \res{\xi=x_k} \frac
{\omega_L(\xi)\omega_M(\xi)\omega_N(\xi)}{{\rm d}\Y(\xi)^2{\rm d}\X(\xi)} \res{\zeta=\xi}  \frac {\omega_K(\zeta)\Omega(\zeta, \xi)}{{\rm
    d}\X(\zeta)}
\eea
The computation could end here, since we have successfully expressed
the derivatives in terms of residues of known differentials: however
this expression is not obviously symmetric in the exchange of the
indices, whereas it should be since it expresses the fourth
derivatives of the free energy. The expression {\em is symmetric}, but not at first sight.
In the double sum the order of the residues is {\em
  immaterial only for the non-diagonal part}: for the diagonal part of
the sum the residue w.r.t. $\zeta$ must be evaluated first.
The non-diagonal part of the sum is
\be
\sum_{k,\ell:\atop \ell\neq k} \res{\xi=x_k} \res{\zeta=x_\ell}
\frac{\omega_L(\xi)\omega_N(\xi)}
{{\rm d}\Y(\xi){\rm d}\X(\xi)}\Omega(\xi,\zeta)
\frac
   {\omega_M(\zeta)\omega_{K}(\zeta)}{{\rm d}\Y(\zeta){\rm
       d}\X(\zeta)}+ \   (M\leftrightarrow L)+(M\leftrightarrow N)
\ee
where the order of the residues is -as we said- immaterial because
they are taken at different points. This term corresponds
diagrammatically to\\

\parbox[c]{3cm}{
\xymatrix{
\lnode{L}\edge{dr} &              &       & \lnode{K}\\
                   & \lnode{$\zeta$}\edge{r}^{\Omega}\edge{dl}&\lnode{$\xi$}\edge{ur}\edge{dr}\\
\lnode{M}          &  & & \lnode{N}
}} +
\parbox[c]{3cm}{
 \xymatrix{
\lnode{M}\edge{dr} &              &       & \lnode{K}\\
                   & \lnode{$\zeta$}\edge{r}^{\Omega}\edge{dl}&\lnode{$\xi$}\edge{ur}\edge{dr}\\
\lnode{N}          &  & & \lnode{L}
}}
+
\parbox[c]{3cm}{\xymatrix{
\lnode{N}\edge{dr} &              &       & \lnode{K}\\
                   & \lnode{$\zeta$}\edge{r}^{\Omega}\edge{dl}&\lnode{$\xi$}\edge{ur}\edge{dr}\\
\lnode{L}          &  & & \lnode{M}
}}    \\
and is manifestly symmetric in $K,L,M,N$. The diagonal part is not
manifestly symmetric but in fact we are going to show that it is.
The diagonal part of the sum together with last term
 is made of the following residues
\bea
\res{\xi=x_k}
\frac{\omega_L(\xi)\omega_N(\xi)}
{{\rm d}\Y(\xi){\rm d}\X(\xi)}\res{\zeta=x_k}
\frac
   {\omega_M(\zeta)\omega_{K}(\zeta)\Omega(\xi,\zeta)}{{\rm d}\Y(\zeta){\rm
       d}\X(\zeta)} + (M\mapsto N\mapsto L)  +   \res{\xi=x_k}
\frac
{\omega_L(\xi)\omega_M(\xi)\omega_N(\xi)}{{\rm d}\Y(\xi)^2{\rm
    d}\X(\xi)} \res{\zeta=\xi}  \frac {\omega_K(\zeta)\Omega(\zeta,
  \xi)}{{\rm
    d}\X(\zeta)}\label{sym}
\eea
where we stress that the residues w.r.t. $\zeta$ have to be evaluated
{\em first}.
For instance, a rather long computation in the local coordinate
$z=\sqrt{\X-\X(x_k)}$ gives
\be
\frac 1 2
\frac{ L\,M\,N\,K'' +  L\,M\,N''\,K+  L\,M''\,N\,K+  L''\,M\,N\,K
+ L\,M\,N\,K\,S_B}{(\Y')^2} - \frac 1 2  \frac {K\,L\,M\,N\,\Y'''}{(\Y')^3}
\ee
where the short-hand notation is as follows
\bea
\omega_L = L(z){\rm d}z\ ,\ \omega_K = K(z){\rm d}z\ ,\ \omega_M =
M(z){\rm d}z\ ,\ \omega_N = N(z){\rm d}z \\
\Omega(z,z') = \le(\frac 1{(z-z')^2} + \frac 1  6  S_B(z,z')\ri){\rm
  d}z{\rm d}z' \ ,
\eea
and $S_B(z,z)$ is the {\em projective connection}, and all quantities
are evaluated at $z=0$.
\subsubsection{$4$-point correlator}
This is the formal expression for
\be
R^{(4)}_{4,0}(q_1,q_2,q_3,q_4) := \frac{\delta^4\mathcal F}{\delta V_1(q_1)V_1(q_2)V_1(q_3)V_1(q_4)}\ ,
\ee
where the formal operator $\delta/\delta V_1(q)$ is defined by
\be
\frac{\delta}{\delta V_1(q)} = \sum_{K=1}^{\infty} q^{-K-1} K \frac
\pa{\pa u_{K,\infty}}\ .
\ee
By summing the four indices of the above derivatives (at least
formally) we obtain
\be
R^{(4)}_{4,0}(q_1,q_2,q_3,q_4) {\rm d}q_1  {\rm d}q_2  {\rm d}q_3  {\rm
  d}q_4 = \Omega^{(4)}_{\X\X}(\zeta(q_1),\zeta(q_2),\zeta(q_3),\zeta(q_4))
\ee
where $\zeta(q)$ is the solution of $\X(\zeta)=q$ on the physical sheet
of the cover $\X:\Sigma_g\to \C P^1$ and
\bea
\Omega^{(4)}_{\X\X}(1,2,3,4) & =&  \sum_{r}\res{\xi=x_k}
  \frac{\Omega(1,\xi)\Omega(2,\xi)\Omega(3,\xi)}{{\rm
  d}\Y^2(\xi){\rm d}\X(\xi)} \res{\zeta=\xi}\Omega(\zeta,\xi)\frac{\Omega(\zeta,4)}{{\rm
    d}\X(\zeta)}+ \\
&+& \sum_{r}\res{\xi=x_k}\sum_k \res{\zeta=x_k}
  \frac{\Omega(1,\zeta)\Omega(4,\zeta)}{{\rm d}\Y(\zeta){\rm
  d}\X(\zeta)}\Omega(\zeta,\xi)\frac {\Omega(2,\xi)\Omega(3,\xi)}{{\rm
  d}\Y(\xi){\rm d}\X(\xi)} +(1\leftrightarrow 2)+(1\leftrightarrow 3)\ .
\eea
Note that this kernel is symmetric in the four variables although not
at first sight, but by the same considerations as before.
In a similar way one can obtain the other four point correlator
\be
R^{(4)}_{0,4}(p_1,p_2,p_3,p_4) := \frac{\delta^4\mathcal F}{\delta V_2(p_1)V_2(p_2)V_2(p_3)V_2(p_4)}\ ,
\ee
where $\delta/\delta V_2(p)$ is defined similarly as before by
\be
\frac \delta{\delta V_2(p)} := \sum_{J=1}^\infty Jp^{-J-1}\frac
\pa{\pa v_J}\ .
\ee
The derivation of the formula is completely parallel hence we only
give the final result
\be
R^{(4)}_{0,4}(p_1,p_2,p_3,p_4) {\rm d}p_1  {\rm d}p_2  {\rm d}p_3  {\rm
  d}p_4 = \Omega^{(4)}_{\Y\Y}(\xi(p_1),\xi(p_2),\xi(p_3),\xi(p_4))
\ee
where $\xi(p)$ is the solution of $\Y(\xi)=p$ on the physical sheet
of the cover $\Y:\Sigma_g\to \C P^1$ and
\bea
\Omega^{(4)}_{\Y\Y}(1,2,3,4) & =&  \sum_{\ell}\res{\xi=y_\ell}
  \frac{\Omega(1,\xi)\Omega(2,\xi)\Omega(3,\xi)}{{\rm
  d}\Y(\xi){\rm d}\X^2(\xi)} \res{\zeta=\xi}\Omega(\zeta,\xi)\frac{\Omega(\zeta,4)}{{\rm
    d}\Y(\zeta)}+ \\
&+& \sum_{\ell}\res{\xi=y_\ell}\sum_s \res{\zeta=y_s}
  \frac{\Omega(1,\zeta)\Omega(4,\zeta)}{{\rm d}\Y(\zeta){\rm
  d}\X(\zeta)}\Omega(\zeta,\xi)\frac {\Omega(2,\xi)\Omega(3,\xi)}{{\rm
  d}\Y(\xi){\rm d}\X(\xi)} +(1\leftrightarrow 2)+(1\leftrightarrow 3)\ .
\eea

\subsection{``Mixed'' fourth order derivatives}
As a further example we compute the derivatives
w.r.t. $u_L,u_M,v_N,v_K$: we leave the derivative w.r.t. $v_K$ last
and perform it at $\X$-fixed
\bea
&& \mathcal F_{LM\tilde N\tilde K} = \pa_{\tilde K} \mathcal F_{LM\tilde
  N} =  \pa_{\tilde K} \sum \res{x_k} \frac{\omega_L\omega_M\omega_{\t
    N}}{{\rm d}\Y{\rm d}\X} =\\
&&=
  \sum \res{x_k} \le(\frac{
(\pa_{\t K}\omega_L)_\X \omega_M\omega_{\t N} +
\omega_L(\pa_{\t K} \omega_M)_\X\omega_{\t N} +
\omega_M \omega_L(\pa_{\t K}\omega_{\t N})_\Y +
\omega_M \omega_L{\rm d}\le(\frac {\omega_{\t N}\omega_{\t K}}{{\rm
    d}\Y{\rm d}\X}\ri)
}
{{\rm d}\Y{\rm d}\X} - \frac{\omega_L\omega_M\omega_{\t
    N}}{{\rm d}\Y{\rm d}\X} \frac{{\rm d}(\pa_{\t K}\Y)_\X}{{\rm d}\Y}\ri)=\\
&&=
 \sum \res{x_k} \le[\frac{
(\pa_{\t K}\omega_L)_\X \omega_M\omega_{\t N} +
\omega_L(\pa_{\t K} \omega_M)_\X\omega_{\t N} +
\omega_M \omega_L(\pa_{\t K}\omega_{\t N})_\Y}
{{\rm d}\Y{\rm d}\X} + \frac{
\omega_M \omega_L}{{\rm d}\Y{\rm d}\X}
{\rm d}\le(\frac {\omega_{\t N}\omega_{\t K}}{{\rm
    d}\Y{\rm d}\X}\ri)
 - \frac{\omega_L\omega_M\omega_{\t
    N}}{{\rm d}\Y^2{\rm d}\X} {\rm d}\le(\frac {\omega_{\t K}}{{\rm
       d}\X}\ri)\ri]=
\eea
\bea
&=&
 \sum \res{\zeta=x_k}
\frac{ \omega_M(\zeta)\omega_{L}(\zeta)}
{{\rm d}\Y(\zeta){\rm d}\X(\zeta)}
\sum \res{\xi=y_\ell}
 \Omega(\xi,\zeta)
\frac {\omega_{\t N} (\xi)\omega_{\t
    K}(\xi)}{{\rm d}\Y(\xi){\rm d}\X(\xi)}+\\
&&-
\sum \res{\zeta=x_k} \le(\frac{ \omega_M(\zeta)\omega_{\t N}(\zeta)}
{{\rm d}\Y(\zeta){\rm d}\X(\zeta)}
\sum \res{\xi=x_\ell}\frac {\omega_L(\xi)\omega_{\t
    K}(\xi)\Omega(\xi,\zeta)}{{\rm d}\Y(\xi){\rm d}\X(\xi)}
+(L\leftrightarrow M)\ri)+\\
&&+
 \sum \res{\zeta=x_k}\le[  \frac{ \omega_M(\zeta)\omega_{L}(\zeta)}
{{\rm d}\Y(\zeta){\rm d}\X(\zeta)}{\rm d}\le( \frac {\omega_{\t N} (\zeta)\omega_{\t
    K}(\zeta)}{{\rm d}\Y(\zeta){\rm d}\X(\zeta)}\ri)
- \frac{\omega_L(\zeta)\omega_M(\zeta)\omega_{\t
    N}(\zeta)}{{\rm d}\Y(\zeta)^2{\rm d}\X(\zeta)} {\rm d}\le(\frac {\omega_{\t K}(\zeta)}{{\rm
       d}\X(\zeta)}\ri)\ri] \ .
\eea
Note that in order to compute effectively the derivative of
$\omega_{\wt K}$ at $\X$-fixed we have used Lemma \ref{lemma2}.
Once more one can check that the resulting expression is symmetric in
$\t N\leftrightarrow \t K$ and $M\leftrightarrow L$. The only terms
which do not have this symmetry at first sight are  the diagonal part of
the double sum over the $x_k$ together with the last term:
\bea
 \mathcal F_{LM\tilde N\tilde K} &=& \sum \res{\zeta=x_k}  \sum \res{\xi=y_\ell}
\frac{ \omega_M(\zeta)\omega_{L}(\zeta)}
{{\rm d}\Y(\zeta){\rm d}\X(\zeta)}
 \Omega(\xi,\zeta)
\frac {\omega_{\t N} (\xi)\omega_{\t
    K}(\xi)}{{\rm d}\Y(\xi){\rm d}\X(\xi)}+\\
&& -
\sum_{k,\ell:\atop \ell\neq k}  \res{\zeta=x_\ell} \res{\xi=x_k} \le(\frac{ \omega_M(\zeta)\omega_{\t N}(\zeta)}
{{\rm d}\Y(\zeta){\rm d}\X(\zeta)}\Omega(\xi,\zeta)
\frac {\omega_L(\xi)\omega_{\t
    K}(\xi)}{{\rm d}\Y(\xi){\rm d}\X(\xi)}
+(L\leftrightarrow M)\ri)+\\
&& +
 \sum \res{\zeta=x_k}  \frac{ \omega_M(\zeta)\omega_{L}(\zeta)}
{{\rm d}\Y(\zeta){\rm d}\X(\zeta)}{\rm d}\le( \frac {\omega_{\t N} (\zeta)\omega_{\t
    K}(\zeta)}{{\rm d}\Y(\zeta){\rm d}\X(\zeta)}\ri)+\\
&&-
\sum_{k}  \res{\zeta=x_k}  \le(\res{\xi=x_k}\frac{ \omega_M(\zeta)\omega_{\t N}(\zeta)}
{{\rm d}\Y(\zeta){\rm d}\X(\zeta)}\Omega(\xi,\zeta)
\frac {\omega_L(\xi)\omega_{\t
    K}(\xi)}{{\rm d}\Y(\xi){\rm d}\X(\xi)}
+ \frac{\omega_L(\zeta)\omega_M(\zeta)\omega_{\t
    N}(\zeta)}{{\rm d}\Y(\zeta)^2{\rm d}\X(\zeta)} {\rm d}\le(\frac {\omega_{\t K}(\zeta)}{{\rm
       d}\X(\zeta)}\ri)\ri)\label{fnclo}
\eea
The same considerations about symmetry done previously apply to the
sum on line (\ref{fnclo}) as well.
\subsubsection{$4$-point correlator}
Using the above computation we can compute the following four-point
correlator
\be
R^{(4)}_{2,2}(q_1,q_2,p_1,p_2) := \frac{\delta^4\mathcal F}{\delta
  V_1(q_1)V_1(q_2)V_2(p_1)V_2(p_2)}\ .
\ee
Performing the multiple summation we find
\be
R^{(4)}_{2,2}(q_1,q_2,p_1,p_2){\rm d}q_1{\rm d}q_2 {\rm d}p_1 {\rm
  d}p_2 = \Omega^{(4)}_{\X\Y}(\zeta(q_1),\zeta(q_2),\xi(p_1),\xi(p_2))
\ee
where $\xi(p)$ is the solution on the physical sheet of $\Y(\xi)=p$ and
\bea
&\Omega^{(4)}_{\X\Y}&\hspace{-10pt}(1,2,\wt 1,\wt 2) := \sum
\res{\zeta=x_k}\sum\res{\xi=y_\ell} \frac
    {\Omega(1,\zeta)\Omega(2,\zeta)}{{\rm d}\Y(\zeta){\rm d}\X(\zeta)}
    \Omega(\zeta,\xi) \frac{\Omega(\wt 1,\xi)\Omega(\wt 2,\xi)}{{\rm
    d}\Y(\xi){\rm d}\X(\xi)} +\\
&& -\sum_{k,r\atop k\neq r} \res{\zeta=x_r}\res{\xi=x_k}  \frac
    {\Omega(1,\zeta)\Omega(\wt 1,\zeta)}{{\rm d}\Y(\zeta){\rm d}\X(\zeta)}
    \Omega(\zeta,\xi) \frac{\Omega(2 ,\xi)\Omega(\wt 2,\xi)}{{\rm
    d}\Y(\xi){\rm d}\X(\xi)} - (1\leftrightarrow 2) +\\
&& +\sum \res{\zeta=x_k}  \frac
    {\Omega(1,\zeta)\Omega(2,\zeta)}{{\rm d}\Y(\zeta){\rm d}\X(\zeta)}
   {\rm d}_\zeta  \frac{\Omega(\wt 1,\zeta)\Omega(\wt 2,\zeta)}{{\rm
    d}\Y(\zeta){\rm d}\X(\zeta)} + \\
&& - \sum \res{\zeta=x_k} \le( \res{\xi=x_k}
\frac
    {\Omega(1,\zeta)\Omega(\wt 1,\zeta)}{{\rm d}\Y(\zeta){\rm d}\X(\zeta)}
    \Omega(\zeta,\xi) \frac{\Omega(2 ,\xi)\Omega(\wt 2,\xi)}{{\rm
    d}\Y(\xi){\rm d}\X(\xi)} + \frac {\Omega(1,\zeta)
      \Omega(2,\zeta)\Omega(\wt 1,\zeta)}{{\rm d}\Y^2(\zeta){\rm
    d}\X(\zeta)}{\rm d}_\zeta\le(\frac {\Omega(\wt 2,\zeta)}{{\rm
    d}\X(\zeta)}\ri)
\ri)
\eea
Repeating the derivation from the beginning one can realize that there
is no need of any other kernel for
\be
R^{(4)}_{3,1}(q_1,q_2,q_3,p_2) := \frac{\delta^4\mathcal F}{\delta
  V_1(q_1)V_1(q_2)V_1(p_3)V_2(p_1)}\ ,
\ee
which is given by
\be
R^{(4)}_{3,1}(q_1,q_2,q_3,p_1){\rm d}q_1{\rm d}q_2 {\rm d}q_3 {\rm
  d}p_1 =
\Omega^{(4)}_{\X\X}(\zeta(q_1),\zeta(q_2),\zeta(q_3),\xi(p_1))\ .
\ee
\subsubsection{Summary of all fourth derivatives}
These three kernels are sufficient for us to write all fourth
derivatives compactly as some new {\em residue formul\ae} (note: the
order in which the integral operators appear is to mean that they are
applied to the variable that appear in the corresponding position in
the kernel)
\bea
&& \pa_{u_K}\pa_{u_J}\pa_{v_L}\pa_{v_M} \mathcal F = \mathcal U_K\mathcal
  U_J\mathcal V_L\mathcal V_M \Omega^{(4)}_{\X\Y}\\
&&  \pa_{u_K}\pa_{v_J}{\pa_{t}}^2 \mathcal F = \mathcal U_K\mathcal T
  \mathcal T \mathcal V_J \Omega^{(4)}_{\X\Y}\\
&& \pa_{u_K}\pa_1\pa_2\pa_3 \mathcal F = \mathcal U_K
  \int_{\pa_1}\int_{\pa_2}\int_{\pa_3} \Omega^{(4)}_{\X\X}\\
&& \pa_{v_J}\pa_1\pa_2\pa_3 \mathcal F = \mathcal V_J
  \int_{\pa_1}\int_{\pa_2}\int_{\pa_3} \Omega^{(4)}_{\Y\Y}\\
 && \pa_1\pa_2\pa_3\pa_4 \mathcal F =
  \int_{\pa_1}\int_{\pa_2}\int_{\pa_3} \int_{\pa_4}\Omega^{(4)}_{\Y\Y}=
   \int_{\pa_1}\int_{\pa_2}\int_{\pa_3} \int_{\pa_4}\Omega^{(4)}_{\X\X}
\eea
where the symbols $\pa_j$ here mean derivatives with respect to
variables not included in the previous items of the list.

\subsection{Higher order correlators}
The computation of any derivative of any order is just a matter of
application of the ``rules of calculus'' outlined previously; in this
fashion one could obtain residue formul\ae\ for any derivative and
possibly develop some diagrammatic rules to help in the
computation. We leave this exercise to the reader who may need it for
his/her application to a specific problem.
The formal ``puncture'' operators 
\be
{\rm d}\X(\xi)\frac {\delta}{\delta V_1(\X(\xi))}\ ,\qquad
{\rm d}\Y(\xi)\frac {\delta}{\delta V_2(\Y(\xi))}
\ee
act as follows on each term
\bea
{\rm d}\X(1)\frac {\delta \Omega(2,3)}{\delta V_1(\X(1))} & =& \sum
\res{\xi=x_k} \frac {\Omega(1,\xi)\Omega(2,\xi)\Omega(3,\xi)}{{\rm
    d}\Y(\xi){\rm d}\X(\xi)}\\
{\rm d}\X(1)\frac {\delta}{\delta V_1(\X(1))}\le( \frac 1 {{\rm
    d}\Y(2)}\ri) &= &\frac 1{{\rm d}\Y^2(2)} {\rm d}_2\le( \frac
  {\Omega(1,2)}{{\rm d}\X(2)}\ri) =\frac 1{{\rm
      d}\Y^2(2)}\res{\xi=2}\frac{\Omega(2,\xi)\Omega(\xi,1)}{{\rm
      d}\X(\xi)}\\
{\rm d}\Y(1)\frac {\delta \Omega(2,3)}{\delta V_2(\Y(1))} & =& -\sum
\res{\xi=y_k} \frac {\Omega(1,\xi)\Omega(2,\xi)\Omega(3,\xi)}{{\rm
    d}\Y(\xi){\rm d}\X(\xi)}\\
{\rm d}\Y(1)\frac {\delta}{\delta V_2(\Y(1))}\le( \frac 1 {{\rm
    d}\X(2)}\ri) &= &-\frac 1{{\rm d}\Y^2(2)} {\rm d}_2\le( \frac
  {\Omega(1,2)}{{\rm d}\Y(2)}\ri) = -\frac 1{{\rm
      d}\X^2(2)}\res{\xi=2}\frac{\Omega(2,\xi)\Omega(\xi,1)}{{\rm
      d}\Y(\xi)}
\eea
Combining these ``rules'' it is easy to obtain any correlator: the
resulting expression {\em will} be symmetric in the exchange of the
variables, although to recognize this some careful analysis of the
residues is required.
\subsection{The Equilibrium correlators}
The derivation of the multiple derivatives of the equilibrium free
energy $\mathcal G$ follows the same lines and the results are the
same formul\ae\ with $\Omega$ replaced by $\wt \Omega$ (clearly there
are no derivatives w.r.t. the filling fractions $\epsilon_j$ which are
now dependent functions). In general the rules of calculus for
$\mathcal G$ are the same as the rule of calculus for $\mathcal F$
with all the instances of the Bergman kernel replaced by the dual
kernel $\wt\Omega$.
\appendix

\section{Explicit form of the regularized integrals}
\label{regularizedintegral}
In this section we provide explicit formul\ae\ for the regularized
integrals used in the definition of the Free energy and the
$\tau$ function of the previous section.\par
 The main tools are   the following properties which were used in the
 proof of the derivatives of the free energy
\bea
  \Y{\rm d}\X &=& \sum_{\alpha=0} \sum_{K=0}^{d_{1,\alpha}} u_{K,\alpha}
\mathcal U_{K,\alpha}\le(\Omega \ri) +   \sum_{\alpha=0}
\sum_{J=0}^{d_{2,\alpha}} v_{J,\alpha} \mathcal V_{J,\alpha}
\le(\Omega \ri) + t\mathcal
  T\le(\Omega \ri)    + \sum_{j=1}^g \epsilon_j
  \mathcal E_j\le(\Omega \ri) -\sum_{\zeta\in \mathcal D_\Y}
  \res{\zeta} \X\Y\Omega \label{master1}\\
 \X{\rm d}\Y &=& -\sum_{\alpha=0} \sum_{K=0}^{d_{1,\alpha}}
u_{K,\alpha} \mathcal U_{K,\alpha}\le(\Omega \ri) -
\sum_{\alpha=0} \sum_{J=0}^{d_{2,\alpha}} v_{J,\alpha} \mathcal
V_{J,\alpha} \le(\Omega \ri) - t\mathcal
  T\le(\Omega \ri)    - \sum_{j=1}^g \epsilon_j
  \mathcal E_j\le(\Omega \ri) -\sum_{\zeta\in \mathcal D_\X}
  \res{\zeta} \X\Y\Omega\label{master2}
\eea
Let us compute $\slint_{q_\alpha}^{\infty_\X}\Y{\rm d}\X$ according to the original definition of regularization: since
 the operator $\slint^{\infty_\X}_{q_\alpha}$ commutes with the
 integral operators/regularizations in
 (\ref{master1}) we obtain immediately
 \bea
 \slint^{\infty_\X}_{q_\alpha}\Y{\rm d}\X &=&
\sum_{\alpha=0} \sum_{K=0}^{d_{1,\alpha}} u_{K,\alpha}
\mathcal U_{K,\alpha}\le(\int^{\infty_\X}_{q_\alpha}\!\!\!\!\! \Omega \ri)
+   \sum_{\alpha=0}
\sum_{J=0}^{d_{2,\alpha}} v_{J,\alpha} \mathcal V_{J,\alpha}
\le(\int^{\infty_\X}_{q_\alpha}\!\!\!\!\! \Omega \ri)+\cr
&& + t\mathcal
  T\le(\int^{\infty_\X}_{q_\alpha}\!\!\!\!\! \Omega \ri)    + \sum_{j=1}^g \epsilon_j
  \mathcal E_j\le( \int^{\infty_\X}_{q_\alpha}\!\!\!\!\! \Omega \ri) -\sum_{\zeta\in \mathcal D_\Y}
  \res{\zeta} \X\Y\int^{\infty_\X}_{q_\alpha}\!\!\!\!\! \Omega
\eea
The differential $\int_{\infty_\X}^{q_\alpha}\!\!\!\!\! \Omega $ is the unique normalized
differential of the third kind with simple poles at $q_\alpha,\infty_\X$ and residues
--respectively-- $+1,-1$. To simplify formul\ae\ let us define for any two points
 $\xi,\eta$ the following function
\bea
\Lambda_{\xi,\eta}(\zeta):= \exp\le[\int_{\zeta_0}^{\zeta}
\int_{\xi}^\eta\!\!\!\!\Omega\ri]\ ;\qquad
\int_{\xi}^\eta\Omega = \frac {{\rm d}\Lambda_{\xi,\eta}}{\Lambda_{\xi,\eta}}
\eea
This is a multivalued function around the $b$-cycles; on the simply connected domain
obtained by dissection of our surface, $\Lambda_{\xi,\eta}$ has a simple pole at $\xi$
and a simple zero at $\eta$. It is defined up to a multiplicative constant (depending
on the base-point for the outer integration), which however will not affect our result.
With this definition we have ($q_0:=\infty_\X,\ p_0:= \infty_\Y$)
\bea
\pa_{u_{0,\alpha}}\mathcal F =  \slint^{\infty_\X}_{q_\alpha}\Y{\rm d}\X &=&
-\sum_{\wt\alpha=0} \res{q_{\wt\alpha}} V_{1,\wt\alpha}(\X) \frac {{\rm d}\Lambda}\Lambda
+\sum_{\beta=0} \res{q_\beta} \le(V_{2,\beta}(\Y) -\X\Y \ri) \frac {{\rm d}\Lambda}\Lambda +\cr
&&+ \sum_{\wt \alpha\neq \{\alpha,0\}} u_{0,\wt\alpha}
\ln\le(\frac{\gamma_{\infty_\X}} {\Lambda(q_{\wt\alpha})}\ri)
+ u_{0,\alpha}\ln\le(\frac{\gamma_{\infty_\X}}{\gamma_{q_\alpha}}\ri)+\cr
&&+ \sum_{\beta=1} v_{0,\beta}\ln\le(\frac {\Lambda(p_\beta)}{\Lambda(\infty_\Y)}\ri) +
t\ln\le(\frac {\gamma_{\infty_\X}}{\Lambda(\infty_\Y)}\ri) +
 \sum_{j=1}^g \frac{\epsilon_j }{2i\pi}\oint_{b_j} \frac {{\rm
 d}\Lambda}{\Lambda} 
\eea
where we have set $\Lambda := \Lambda_{q_\alpha,\infty_\X}$ and
\bea
\ln \gamma_{\infty_\X} &:=& \lim_{\epsilon\to \infty_\X}
             \ln \le({\Lambda_{q_\alpha,\infty_\X}} \X\ri)\cr
\ln \gamma_{q_\alpha} &:=& \lim_{\epsilon \to q_\alpha}
            \ln \le({\Lambda_{q_\alpha,\infty_\X}} (\X - Q_\alpha)\ri)
\eea
The formul\ae\ for the derivatives w.r.t. $v_{0,\alpha}$ are obtained by interchanging all
the r\^oles of $\X,\infty_\X,q_\alpha$ with $\Y,\infty_\Y,p_\alpha$.\par
Finally the formula for the $t$-derivative
\bea
\pa_t \mathcal F &=& \slint_{\infty_\Y}^{\infty_\X} \Y{\rm
 d}\X-\sum_\alpha v_{0,\alpha} =
-\sum_{\wt\alpha=0} \res{q_{\wt\alpha}} V_{1,\wt\alpha}(\X) \frac {{\rm d}\Lambda}\Lambda
+\sum_{\beta=0} \res{q_\beta} \le(V_{2,\beta}(\Y) -\X\Y \ri) \frac {{\rm d}\Lambda}\Lambda +\cr
&&+ \sum_{\wt \alpha=1} u_{0,\wt\alpha}
\ln\le(\frac{\gamma_{\infty_\X}} {\Lambda(q_{\wt\alpha})}\ri)
+ \sum_{\beta=1} v_{0,\beta}\ln\le(\frac {\Lambda(p_\beta)}{\gamma_{\infty_\Y}}\ri) +
t\ln\le(\frac {\gamma_{\infty_\X}}{\gamma_{\infty_\Y}}\ri) +
 \sum_{j=1}^g \frac{\epsilon_j }{2i\pi}\oint_{b_j} \frac {{\rm
 d}\Lambda}{\Lambda}  + t\label{mu}
\eea
where -this time-
\bea
\Lambda&:=& \Lambda_{\infty_\Y,\infty_\X}\cr
\ln(\gamma_{\infty_\X}) &:=& \lim_{\epsilon\to \infty_\X} \ln\le( {\Lambda}\X\ri)\cr
\ln(\gamma_{\infty_\Y}) &:=& \lim_{\epsilon\to \infty_\Y} \ln\le(
 \frac {\Lambda}\Y \ri)
\eea
The extra $"\sum_\alpha v_{0,\alpha}"$ which cancels with the same
 term in the expression for $\mu$ is due to a careful analysis of the regularization
prescription for the following term in the computation
\bea
\res{\infty_\Y}\X\Y \int_{\infty_\Y}^{\infty_\X}\Omega =
 \lim_{\epsilon\to \infty_\Y} \le(\int_{\epsilon}^{\infty_\X} \res{\infty_\Y}\X\Y \Omega
+\X(\epsilon)\Y(\epsilon)\ri) =
 -t-\sum_\alpha v_{0,\alpha} + \slint_{\infty_\Y}^{\infty_\X}\res{\infty_\Y}\X\Y\Omega\ .
\eea

Note that, in all these formul\ae, the $b$-periods of $\frac{{\rm
    d}\Lambda}{\Lambda}$ are the Abel map of the two poles of this differential.
\section{Example: one cut case (genus $0$) and conformal maps}
The formul\ae\ for the derivatives simplify drastically in case the
curve $\Sigma_g$ is a rational curve. In this case, introducing a
global coordinate $\lambda$ (as explained in \cite{F1, F2}) with a
zero at $\infty_\Y$ and a pole at $\infty_\X$ and suitably normalized
one can always write the two functions $\X,\Y$ as
\bea
&& \X = \gamma\lambda +\sum_{K=0}^{d_{2,\infty}} A_{K,\infty} \lambda^{-K}
+ \sum_\alpha \sum_{K=0}^{d_{2,\alpha}}A_{K,\alpha}(\lambda - \lambda_{\wt\alpha})^{-K-1} + \sum_{j=1}^{s}
\frac {F_{j}}{\lambda - \lambda_{j,Y}} \cr
&& \Y = \frac \gamma \lambda +\sum_{J=0}^{d_{1,\infty}} B_{j,\infty}
\lambda^j +
\sum_\alpha \sum_{J=0}^{d_{1,\alpha}} B_{J,\alpha}(\lambda -
\lambda_{\alpha})^{-J-1} + \sum_{j=1}^r \frac {G_j}{\lambda -\lambda_{j,X}}\label{uniform}\\
&& Q_\alpha:= \X(\lambda_\alpha)\ ,\ \ P_{\wt\alpha}:=\Y(\lambda_{\wt\alpha})\ .
\eea
The parameters $\gamma, A_j,F_j$ and $B_j,G_j,j=1\dots,s$ are not
independent but are constrained  by the following set of linear equations
(in $\gamma,A_j,B_j,F_j,G_j$)
\be
\X'(\lambda_{j,X}) = 0\ ,\ j=1,\dots, r\ ;\qquad
\Y'(\lambda_{j,Y}) = 0\ ,\ j=1,\dots, s\ .
\ee
As we have already mentioned, the equivalent of the Bergman kernel is simply
\be
\Omega(\lambda,\mu) = \frac {{\rm d}\lambda{\rm
    d}\mu}{(\lambda-\mu)^2}\ .
\ee
This is the kernel of the derivative followed by projection to the
principal part.  For example
the differentials $\omega_{K,\infty}$
\bea
\omega_{K,\infty}(\lambda) &=& -\res{\mu=\infty} \frac
      {\X^K(\mu)}K\Omega(\lambda,\mu) =\frac 1 K \le(\X^K\ri)_+ {\rm
    d}\lambda\ ,
\eea
where the $\pm$-subscripts mean the polynomial or Laurent part of the
expression enclosed in the brackets.
Similar completely explicit formul\ae\ for all other differentials
in the lists (\ref{diff1}, \ref{diff2}, \ref{diff3}) are left to the
reader.
\par
The coordinates are given by the usual formul\ae\ (\ref{coord1}).
The free energy can be written in a quite explicit form using the following simplifications
due to the existence of a global coordinate $\lambda$ ($\lambda_0:=\infty,\ \lambda_{\wt 0}:= 0$)
\bea
\pa_{u_{0,\alpha}} \mathcal F &=& -\sum_{\beta\geq 0} \res{\lambda=\lambda_{\beta}}
V_1(\X) \frac {{\rm d}\lambda}{ \lambda - \lambda_{\alpha}}
+ \sum_{\wt\beta\geq 0}\res{\lambda =\lambda_{\wt\beta}}\le(V_2(\Y)-\X\Y\ri)
\frac {{\rm d}\lambda}{\lambda - \lambda_\alpha} +\cr
&&+\sum_{\beta\neq \{\alpha,0\}} u_{0,\beta}
\ln \le(\gamma {(\lambda_\beta-\lambda_\alpha) }\ri) +
 u_{0,\alpha}\ln \le(\frac\gamma{\X'(\lambda_\alpha)} \ri)
 +\sum_{\wt\beta \geq 0} v_{0,\beta} \ln\le(\frac {\lambda_\alpha}
 {\lambda_\alpha-\lambda_{\wt\beta}} \ri)
 + t\ln \le({\lambda_\alpha}\gamma\ri)\\
\pa_{v_{0,\wt\alpha}} \mathcal F &=& 
-\sum_{\wt\beta\geq 0} \res{\lambda=\lambda_{\wt\beta}}
V_{2,\wt\beta} (\Y) \frac {\lambda_{\wt\alpha}{\rm d}\lambda}{ \lambda( \lambda_{\wt\alpha}-\lambda)}
+ \sum_{\beta\geq 0}\res{\lambda =\lambda_{\beta}}\le(V_1(\X)-\X\Y\ri)
 \frac {\lambda_{\wt\alpha}{\rm d}\lambda}{ \lambda( \lambda_{\wt\alpha}-\lambda)} +\cr
&&-\sum_{\wt\beta\neq \{\wt\alpha,0\}} v_{0,\wt\beta}
\ln \le(\frac {(\lambda_{\wt\alpha})^2}{(\lambda_{\wt\beta}-\lambda_{\wt\alpha})\gamma}\ri) -
 v_{0,\wt\alpha}\ln \le(\frac {(\lambda_{\wt\alpha})^2\Y'(\lambda_{\wt\alpha})}\gamma\ri)
 +\sum_{\beta \geq 0} u_{0,\beta} \ln\le(\frac {\lambda_\beta}
 {\lambda_{\beta}-\lambda_{\wt\alpha}} \ri)
 + t\ln \le(\frac   \gamma{\lambda_{\wt\alpha}}\ri)
\eea
since $\Lambda_{q_\alpha,\infty_\X} = \frac 1{\lambda - \lambda_\alpha}$ and
$\Lambda_{p_{\wt\alpha},\infty_\Y} = \frac \lambda {\lambda - \lambda_{\wt\alpha}}$
Moreover, using this time $\Lambda_{\infty_\X,\infty_\Y} = \lambda$ and formula (\ref{mu})
\bea
\pa_t \mathcal F &=&\sum_{\beta\geq 0} \res{\lambda=\lambda_{\beta}}
V_{1,\beta}(\X) \frac {{\rm d}\lambda}{ \lambda}
- \sum_{\wt\beta\geq 0}\res{\lambda =\lambda_{\wt\beta}}\le(V_{2,\wt\beta}(\Y)-\X\Y\ri)
\frac {{\rm d}\lambda}{\lambda } +\cr
&&+\sum_{\beta=1} u_{0,\beta}
\ln \le({\lambda_\beta}\gamma\ri)
 -\sum_{\wt\beta = 1 } v_{0,\beta} \ln\le(\frac {\lambda_\beta}\gamma\ri)
 + t\ln \le(\gamma^2\ri) +t
\eea
By computing the other residues one can get explicit formulas for the
Free energy  in terms of
the uniformization (\ref{uniform}) and using Thm. \ref{main}.\par
Denoting as before by $x_k$ and $y_\ell$ the critical points of the
functions $\X$ and $\Y$\footnote{Note that the set of points
  $\{\lambda_{j,X}\}$ is a subset of the $\{x_k\}$ and similarly for
  the $\{\lambda_{j,Y}\}$ and $\{y_k\}$.}  respectively we have as example of fourth
point correlators
\bea
&&-R^{(4)}_{4,0}(\mu_1,\mu_2,\mu_3,\mu_4)\X'(\mu_1)\X'(\mu_2)\X'(\mu_2)\X'(\mu_4)
=\cr
&&=\sum_{k,r\atop k\neq r}
 \frac {(\mu_1-x_k)^{-2}(\mu_2-x_k)^{-2}}{\Y'(x_k)\X''(x_k)} \frac{1}{(x_k-x_r)^2} \frac
       {(\mu_3-x_r)^{-2}(\mu_4-x_r)^{-2}}{\Y'(x_r)\X''(x_r)} + (\mu_1\leftrightarrow \mu_3) +
       (\mu_1\leftrightarrow \mu_4) +\\
&&+ \sum_{k}
\frac 1 {6 \Y''' (\X'')^4}
\bigg[
\frac{
  \le(2 \Y' (\X''')^2 + 2 \Y'' \X''\X'''-3(\X'')^2 \Y''' -3\X^{(iv)}\X''\Y'\ri)
}
{(\mu_1-x_k)^{2} (\mu_2-x_k)^{2} (\mu_3-x_k)^{2}(\mu_4-x_k)^{2}}
+\\
&& +
\frac{18(\X'')^2\Y'\le((\mu_1-x_k)^{-2} +\hbox{cyc}\ri)}
{(\mu_1-x_k)^{2} (\mu_2-x_k)^{2} (\mu_3-x_k)^{2}(\mu_4-x_k)^{2}}
 -\frac {2\X'''\X''\Y'
\le((\mu_1-x_k)^{-1}+\hbox{cyc}\ri) }
{(\mu_1-x_k)^{2} (\mu_2-x_k)^{2} (\mu_3-x_k)^{2}(\mu_4-x_k)^{2}}
\bigg]\Bigg|_{\lambda=x_k}\ .
\eea
Here the expression looks more complicated than necessary because the
derivatives are taken w.r.t. $\lambda$.

Higher order correlators are of increasingly cumbersome expression but
in principle they are easily computed using the general calculus
outlined in the main text.
\subsection{Conformal maps}
A further simplification of the formul\ae\ arises in the case the
functions $\Y$ and $\X$ above describe the Riemann uniformization  and
its Schwartz reflected of a  simply-connected domain $\mathcal D$ in
the $\X$ plane \cite{}. We recall that all our formulas can be easily
adapted to the description of simply and multiply connected domains
(the number of connected components being the genus of the curve) by
taking the curve $\Sigma_g$ as an $M$-curve in the sense of Harnack \cite{harnack}:
namely a curve with an anti-holomorphic involution
$\varphi:\Sigma_g\to\Sigma_g$  having $g+1$
contours of fixed points and such that
\be
\X(\zeta) = \overline{\Y(\varphi(\zeta))}\ .
\ee
In genus zero and with the normalization used in the previous
paragraph for the uniformizing coordinate, the anti-holomorphic involution would be $\lambda\to
\frac 1 {\overline \lambda}$. The two functions $\Y$ and $\X$ then
satisfy
\be
\X(\lambda) = \overline{\Y\le(\frac 1 {\overline \lambda}\ri)}
\ee
Since $\X(\lambda)$ is now the uniformizing map of a simply connected
domain $\mathcal D$ it follows from the general properties of such
maps that $\X$ maps biholomorphically the outer region $\C\setminus
\mathcal D$ to the outside of the unit disk in the
$\lambda$-plane. This means that the zeroes of ${\rm d}\X$ all lie
inside the unit disk and hence the zeroes of ${\rm d}\Y$ (which is the
Schwartz function of the domain) all lie outside.\\
The Free energy of the two-matrix model under this reduction
$v_K=\overline u_K$, reduces to the tau-function of Jordan curves
studied by Zabrodin at al. \cite{KKWZ, KMZ, MWZ1, WZ, WZ2, Zab} as explained in \cite{F1, F2}
\be
\mathcal F = \frac 1{4\pi^2}\int_{\mathcal D}\int_{\mathcal D}{\rm
  d}^2 \X{\rm d}^2 \wt \X
\ln\le| \frac 1 \X-\frac 1{\wt \X}\ri|
\ee
The coordinates $u_K$ are the so--called  {\em exterior harmonic moments} of the
region
\bea
t = \frac 1{2i \pi}\int_{\mathcal D}{\rm d}\X\wedge {\rm d}\ov \X = \ov
t \\
u_k = \frac 1{2i \pi}\int_{\mathcal D} \X^K {\rm d}\X\wedge {\rm d}\ov
\X\ .
\eea
The Free energy is in this case a real analytic function of the
harmonic moments $\mathcal F =(u_K,\ov{u_K},t)$ and the previous
formul\ae\ for the fourth derivatives\footnote{The third derivatives
  were computed in \cite{Zab}.}
can be translated  into contour integral-formul\ae\ which in turn
could be written in terms of the Green's function of the Laplacian for
the given region.
It is also clear that effective formul\ae\ can be obtained for the
multiply connected domains which correspond to
higher genus $M$-curves considered in this context.

\section {An extended moduli space}
\label{extended}
The moduli space considered in this paper could be easily extended in
the spirit of \cite{Kric} by considering instead of functions $\X,\Y$
some normalized second-kind differentials ${\rm d}\X,{\rm d}\Y$: this
generalization has probably no relevance in the context of matrix
models, nevertheless we sketch the main extra features.
The practical difference is that now we may still think of {\em
  multivalued} functions $\X,\Y$ with the properties
\bea
\X(\zeta+b_j) = \X(\zeta) + A_j\\
\Y(\zeta+b_j) = \Y(\zeta) + B_j\ ,
\eea
whereas the functions have no multivaluedness along the $a$-cycles.
The rest of the description of the moduli space is exactly as in
Sect. \ref{setting}. Note that this moduli space is ``larger'' than
the moduli space of \cite{Kric} because we are also considering the
position of some zeroes of our primary differentials.\par
After dissection of the surface $\Sigma_g$ along the chosen cycles
$\{a_j,b_j\}_{j=1,\dots,g}$ and along the fixed contours between the
non hard-edge poles we obtain a simply connected domain over which we
will consider the functions $\X=\int {\rm d}\X,\Y=\int {\rm d}\Y$. In
this domain the same asymptotics as in (\ref{coord1}) are valid (where the
``potentials'' are discontinuous across the cuts along which we have
dissected the surface).
The free energy (we should probably call it rather
``tau'' function) would be defined by the same formul\ae\
(\ref{firstders}) except for the fact that the
$\epsilon_j$-derivatives should be  replaced by the
formulas below and we should consider the derivatives w.r.t. the extra
moduli $A_j,B_j$
\bea
\mathbb A_j:= \pa_{A_j} \mathcal F &=& \frac 1{2i\pi} \le(\oint_{a_j} \Y\X{\rm d}\Y
-\frac 12 \epsilon_jB_j\ri) \\
\mathbb B_j:= \pa_{B_j} \mathcal F &=& \frac 1{2i\pi}\le( \oint_{a_j} \Y\X{\rm d}\X
+\frac 12\epsilon_jA_j\ri)\\
\Gamma_j:= \pa_{\epsilon_j} \mathcal F &=&   \frac
1{2i\pi}\le(\frac 1 2 A_jB_j - B_j\X(\zeta)+\int_{\zeta}^{\zeta+b_j}\!\!\!\! \Y{\rm
  d}\X\ri)\cr
&=&   -\frac
1{2i\pi}\le(\frac 1 2 A_jB_j - A_j\Y(\zeta)+\int_{\zeta}^{\zeta+b_j}\!\!\!\! \X{\rm
  d}\Y\ri)
\eea
The equivalence of the two last lines is given by integration by
parts. Also,
the  last integrals may seem to depend
on the base-point of integration: in fact they do not as one may check
by computing the differential at $\zeta$.\par
Besides the differentials considered in (\ref{diff1}, \ref{diff2},
\ref{diff3}) one also has
\bea
(\pa_{A_j} \X)_\Y{\rm d}\Y &=& \frac 1{2i\pi}\oint_{a_j} \Y\Omega=:- \mathcal A_j (\Omega)\cr
(\pa_{B_j}\Y)_\X{\rm d}\X &=& \frac 1 {2i\pi}\oint_{a_j} \X\Omega=:\mathcal B_j(\Omega)
\eea
These formul\ae\ are obtained by noticing that $(\pa_{B_j} \Y)_\X {\rm
  d}\X$ is a holomorphic {\em multivalued} differential with monodromy
  only around the corresponding $b$-cycle
\be
(\pa_{B_j} \Y)_\X {\rm
  d}\X\bigg|_{\zeta}^{\zeta+b_k} = -\delta_{jk} {\rm d}\X(\zeta)\ .
\ee
The integral formula has the same properties and hence we have the
  equality. The reasoning for $(\pa_{A_j}\X)_{\Y}{\rm d}\Y$ is
  symmetric. Note that -using the thermodynamic identity- we have
\be
(\pa_{B_j} \X)_\Y{\rm d}\Y = -\frac 1 {2i\pi}\oint_{a_j} \X\Omega\ .
\ee
The considerations to prove the compatibility of the above equations
  are similar to the previous case with one notable exception we want
  to bring to the attention of the reader; in the computations of the
  second derivatives one is lead to considering integrals of the form
\be
\oint_{a_j}\X\oint_{a_k}\Y\Omega\ ,\ \
  \oint_{a_j}\X\oint_{a_k}\X\Omega\ ,\ \
  \oint_{a_j}\Y\oint_{a_k}\Y\Omega \ .
\ee
These integrals do not depend on the order only if $j\neq k$: in fact
  we have
\bea
\oint_{a_j}\X\oint_{a_j}\Y\Omega &=& \oint_{a_j}\Y\oint_{a_j}\X \Omega +
  2i\pi \oint_{a_j} \X{\rm d}\Y\cr
  \oint_{a_j}\X\oint_{a_k}\X\Omega &=&  \oint_{a_j}\X\oint_{a_j}\X \Omega +
  2i\pi \oint_{a_j} \X{\rm d}\X = \oint_{a_j}\X\oint_{a_j}\X \Omega \cr
  \oint_{a_j}\Y\oint_{a_j}\Y\Omega &=&  \oint_{a_j}\Y\oint_{a_j}\Y \Omega +
  2i\pi \oint_{a_j} \Y{\rm d}\Y =  \oint_{a_j}\Y\oint_{a_j}\Y\Omega
\eea
Another kind of integrals that one encounters are of the type
\be
\oint_{a_j}\Y\oint_{b_k} \Omega = 2i\pi\oint_{a_j} \Y \omega_k
\ee
Here one has to use the following rule for exchanging the order of the
  integrals: suppose that a specific choice of the
  homology representatives of $a_j$ and $b_j$ intersect at
  the point $\zeta_0$, then
\bea
\oint_{\zeta\in a_j}F(\zeta)\oint_{\xi\in b_k}\Omega(\zeta,\xi) =
  \oint_{\zeta\in a_j}(F(\zeta)-F(\zeta_0)) \oint_{\xi\in
  b_k}\Omega(\zeta,\xi) +F(\zeta_0) \oint_{\zeta\in a_j}\oint_{\xi\in b_k}\Omega(\zeta,\xi)=\\
= \oint_{\xi\in b_k} \oint_{\zeta\in
  a_j}(F(\zeta)-F(\zeta_0))\Omega(\zeta,\xi)- 2i\pi\delta_{jk}
  F(\zeta_0)  =   2i\pi\delta_{jk} F(\zeta_0)+\oint_{\xi\in b_k}\oint_{\zeta\in a_j}F(\zeta)\Omega(\zeta,\xi)
\eea
By following similar arguments one can prove that
\be
2\mathcal F = 2\mathcal F_0 + \sum_{j=1}^g (A_j\mathbb A_j + B_j\mathbb B_j)
\ee
where $\mathcal F_0$ is given by the same formula (\ref{freeenergy}) (with the
new meaning of $\Gamma_j$, though). The proof rests on the identity
\bea
\Y{\rm d}\X &=& \sum_{\alpha=0} \sum_{K=0}^{d_{1,\alpha}} u_{K,\alpha}
\mathcal U_{K,\alpha}\le(\Omega \ri) +   \sum_{\alpha=0} \sum_{J=0}^{d_{2,\alpha}} v_{J,\alpha}
\mathcal V_{J,\alpha} \le(\Omega \ri) +\\
&&+ t\mathcal
  T\le(\Omega \ri)    + \sum_{j=1}^g \le(\epsilon_j
  \mathcal E_j\le(\Omega \ri) +\frac 1{2i\pi}\oint_{a_j} \X\Omega +
  \frac 1{2i\pi}\oint_{a_j} \Y\Omega\ri) + \sum_{\zeta\in\mathcal
    D_\Y} \res{\zeta} \X\Y\Omega\ ,
\eea
which is proved as before by matching the singular behaviors of both
sides at all possible singularities and by checking that both sides
have the same multivaluedness around the $a$ and  $b$-cycles and the same periods.
\subsection{Higher order derivatives}
In order to write compactly the second derivatives let us denote by
$\pa$ any derivative w.r.t. one of the parameters $u_{K,\alpha},
Q_\alpha, X_j,v_{J,\alpha},
P_\alpha, Y_j$. Beside the second derivatives already computed, the
new ones are given by the formul\ae
\bea
&&
\pa_{A_j}\pa_{A_k}\mathcal F =\mathcal A_j\mathcal A_k \Omega
 \ ,\qquad \pa_{B_j}\pa_{B_k}\mathcal F = \mathcal B_j\mathcal B_k\Omega \cr
&&
\pa_{A_j}\pa_{B_k} \mathcal F = \mathcal A_j\mathcal B_k \Omega + \frac
{\delta_{jk}}{4i\pi} \epsilon_k\cr
&&
\pa_{A_j}\pa_{\epsilon_k} \mathcal F = \mathcal A_j\mathcal E_k \Omega + \frac
{\delta_{jk}}{4i\pi} B_j\cr
&&
\pa_{B_j}\pa_{\epsilon_k} \mathcal F = \mathcal B_j\mathcal E_k \Omega 
- \frac {\delta_{jk}}{4i\pi} A_k\cr
&&
\pa_{B_j}\pa \mathcal F = \mathcal B_j
\int_{\pa} \Omega\ ;\qquad
 \pa_{A_j} \pa \mathcal F = \mathcal A_j
 \int_{\pa} \Omega
\eea
We remark that the order of the integral operators acting on $\Omega$ is relevant because
$\Omega$ is singular on the diagonal: for instance
\be
\mathcal A_j\mathcal B_k \Omega= \mathcal B_k\mathcal A_j \Omega 
-\frac{\delta_{jk}}{2i\pi} \epsilon_k\ .
\ee
In order to compute all higher derivatives and loop correlators we need to
specify the relevant additional Rauch variational formul\ae: besides
those considered in (\ref{rauchforms}) we need the ones related to the
extra moduli
\bea
(\pa_{A_j}\Omega)_\X (1,2)&=&-\frac 1{2i\pi}\oint_{\xi\in a_j}
\le(\Y(\xi)\Omega^{(3)}_\X (1,2,\xi) - \frac
   {\Omega(1,\xi)\Omega(2,\xi)}{{\rm d}\X(\xi)}\ri)\cr
(\pa_{B_j}\Omega)_\X (1,2)&=&\frac 1{2i\pi}\oint_{\xi\in a_j}
\X(\xi)\Omega^{(3)}_\X (1,2,\xi) \cr
(\pa_{B_j}\Omega)_\Y (1,2)&=&\frac 1{2i\pi}\oint_{\xi\in a_j}
\le(\X(\xi)\Omega^{(3)}_\Y (1,2,\xi) + \frac
   {\Omega(1,\xi)\Omega(2,\xi)}{{\rm d}\Y(\xi)}\ri)\cr
(\pa_{A_j}\Omega)_\Y (1,2)&=&-\frac 1{2i\pi}\oint_{\xi\in a_j}
\Y(\xi)\Omega^{(3)}_\Y (1,2,\xi)\label{rauchadd}
\eea
We briefly justify these formul\ae. Suppose $\omega$ is any of our differentials and consider the
function $\omega/{\rm d}\X$ (or symmetric argument for $\Y$). This function has poles at the zeroes
of ${\rm d}\X$ and possibly constant monodromy around a $b$-cycle. Thinking of it as a function 
of $\X$ the monodromy condition reads ($c$ is $0$ or $1$ depending on the case chosen, but the 
argument is unaffected)
\be
\frac{\omega}{{\rm d}\X}(\X + A_j) - \frac {\omega}{{\rm d}\X}(\X) = c
\ee
Taking the derivative w.r.t. $A_j$ at $\X$-fixed we have
\be
(\pa_{A_j}\omega)_\X\bigg|_{\zeta}^{\zeta+b_j} = {\rm d}\le(\frac {\omega}{{\rm d}\X} \ri)
\ee
Considering with some care the singularities at the zeroes of ${\rm d}\X$ and this multivaluedness 
one gets
\be
(\pa_{A_j}\omega)(\zeta) = -\sum \res{\xi=x_k} \frac {\omega(\xi)\Omega(\xi,\zeta)
\mathcal A_j(\Omega)(\xi)}{{\rm d}\X(\xi){\rm d}\Y(\xi)} + \frac 1 {2i\pi}\oint_{a_j} 
\frac {\Omega(\zeta,\xi)\omega(\xi)}{{\rm d}\X(\xi)}\ .
\ee
This gives the previous extended Rauch formul\ae.\par
Using these expressions for the variation of the Bergman kernel one can obtain all third 
derivatives. Besides those already considered in (\ref{third1}, \ref{third2}, \ref{third3}) we
find also
\bea
\pa_{A_j}\pa_{A_k}\pa_{A_\ell}\mathcal F = \mathcal A_j\mathcal A_k \mathcal A_\ell \Omega^{(3)}_\Y
\ ;\qquad
\pa_{B_j}\pa_{B_k}\pa_{B_\ell}\mathcal F = \mathcal B_j\mathcal B_k \mathcal B_\ell \Omega^{(3)}_\X
\cr
\pa_{A_j}\pa_{A_k}\pa \mathcal F = \int_\pa\mathcal A_j\mathcal A_k\Omega^{(3)}_\Y\ ;\qquad
\pa_{B_j}\pa_{B_k}\pa \mathcal F = \int_\pa\mathcal B_j\mathcal B_k\Omega^{(3)}_\X\cr
\pa_{A_j}\pa_{B_k}\pa_{\epsilon_\ell} \mathcal F = 
\mathcal A_j\mathcal B_k\mathcal E_\ell \Omega^{(3)}_\X +
\frac 1{2i\pi}\oint_{a_j} \frac {\mathcal B_k(\Omega)\mathcal E_j(\Omega)}{{\rm d}\X} 
-\frac {\delta_{jk}\delta_{kl}}{4i\pi}\cr
\pa_{A_j}\pa_{B_k}\pa_{v_{J,\alpha}}\mathcal F = 
\mathcal A_j\mathcal B_k\mathcal V_{J,\alpha} \Omega^{(3)}_\Y + \frac 1{2i\pi}\oint_{a_k} 
\frac{\mathcal A_j(\Omega)\mathcal V_{J,\alpha}(\Omega)}{{\rm d}\Y}\cr
\pa_{A_j}\pa_{B_k}\pa_{u_{K,\alpha}}\mathcal F = 
\mathcal A_j\mathcal B_k\mathcal U_{K,\alpha} \Omega^{(3)}_\X + \frac 1{2i\pi}\oint_{a_j} 
\frac{\mathcal B_k(\Omega)\mathcal U_{K,\alpha}(\Omega)}{{\rm d}\X}\cr
\pa_{A_j}\pa_{u_{K,\alpha}}\pa_{v_{J,\beta}}\mathcal F = 
\mathcal A_j \mathcal U_{K,\alpha} \mathcal  V_{J,\beta} \Omega^{(3)}_{\Y}\ ;
\qquad \pa_{B_j}\pa_{u_{K,\alpha}}\pa_{v_{J,\beta}}\mathcal F = 
\mathcal B_j \mathcal U_{K,\alpha} \mathcal  V_{J,\beta} \Omega^{(3)}_{\X}\cr
\pa_{A_j}\pa_{u_{K,\alpha}}\pa_{u_{J,\beta}}\mathcal F = 
\mathcal A_j \mathcal U_{K,\alpha} \mathcal  U_{J,\beta} \Omega^{(3)}_{\X} 
+\frac 1{2i\pi}\oint_{a_j}\frac { \mathcal U_{K,\alpha} (\Omega)
\mathcal  U_{J,\beta}(\Omega)}{{\rm d}\X}\cr
\pa_{B_j}\pa_{v_{K,\alpha}}\pa_{v_{J,\beta}}\mathcal F = 
\mathcal B_j \mathcal V_{K,\alpha} \mathcal  V_{J,\beta} \Omega^{(3)}_{\Y} 
+\frac 1{2i\pi}\oint_{a_j}\frac { \mathcal V_{K,\alpha} (\Omega)
\mathcal  V_{J,\beta}(\Omega)}{{\rm d}\Y}\cr
\pa_{A_j}\pa_{v_{K,\alpha}}\pa_{v_{J,\beta}}\mathcal F = 
\mathcal A_j \mathcal V_{K,\alpha} \mathcal  V_{J,\beta} \Omega^{(3)}_{\Y}\ ;\qquad
\pa_{B_j}\pa_{u_{K,\alpha}}\pa_{u_{J,\beta}}\mathcal F = 
\mathcal B_j \mathcal U_{K,\alpha} \mathcal  U_{J,\beta} \Omega^{(3)}_{\X}
\cr
\pa_{A_j} \pa_t\pa \mathcal F = \int_{\pa}\mathcal T\mathcal A_j \Omega^{(3)}_\Y\ ;\qquad
\pa_{B_j} \pa_t\pa \mathcal F = \int_{\pa}\mathcal T\mathcal B_j \Omega^{(3)}_\X\ .
\eea
\subsubsection{Order four and higher}
It is clear that the formul\ae\ become rather long due to many case-distinctions. However the 
reader should be able to compute any derivative of order four or higher by using the same 
rules of calculus outlined in the main text, with the additional Rauch formul\ae\ 
(\ref{rauchadd}).

\section{General definition of regularized integrals}
\label{genreg}
Let $\omega$ be a meromorphic differential with poles at the points 
$\zeta_\rho$, $\rho = 0,\dots $.
 Let $z_\rho$ be chosen and fixed local parameters at $\zeta_\rho$. 
 Let $\omega_j$ be the Abelian differentials of the first kind normalized w.r.t. the $a$-cycles 
 of a given choice of basis $\{a_j,b_j\}$ in the homology of the curve.
 Then we have
 \be
 \omega = \sum_{\rho\geq 0}\sum_{K\geq 1} \frac 1 K \res{\zeta_\rho} (z_\rho)^{K}
 \res{\zeta_\rho} (z_\rho)^{-K} \Omega +\sum_{\rho\geq 1} \le(\res{\zeta_\rho}\omega\ri) 
 \int_{\zeta_0}^{\zeta_\rho} \Omega + \sum_{j=1}^g \le(\oint_{a_j}\omega\ri)\omega_j
 \label{genmaster}
 \ee
 The regularized integral from $\xi$ to $\eta$ is defined for a homology class of contours in the
 punctured surface: in general one has to dissect the surface along the $a,b$-cycles and along a 
 set of mutually non-intersecting segments joining the poles of $\omega$ in such a was as to have
 a simply connected domain. Choosing an arbitrary path within this simply connected region
 and joining the two chosen points we have (supposing that both $\xi,\eta$ are poles of $\omega$)
 \bea
 \slint_{\xi}^\eta \omega &=&
  \sum_{\rho\geq 0}\sum_{K\geq 1} \frac 1 K \res{\zeta_\rho} (z_\rho)^{K}
 \res{\zeta_\rho} (z_\rho)^{-K} \frac {{\rm d}\Lambda}\Lambda 
  +
  \sum_{\rho\geq 0\atop
  \zeta_\rho\not\in\{\xi,\eta\} } \le(\res{\zeta_\rho}\omega\ri) 
 \ln\le(\frac {\Lambda(\zeta_\rho)}{\gamma_\xi} \ri)+\cr
 &&+ \le(\res{\eta}\omega\ri)\ln\le(\frac {\gamma_\eta}{\gamma_\xi}\ri)+  
 \sum_{j=1}^g \le(\oint_{a_j}\omega\ri)\oint_{b_j}  \frac {{\rm d}\Lambda}\Lambda
 \label{slint}\\
 \Lambda&:=& \exp\le(\int    \int_\xi^\eta \Omega\ri):\ \ 
 \frac {{\rm d}\Lambda}\Lambda:= \int_{\xi}^\eta\Omega \cr
 \gamma_\xi&:=& \lim_{\epsilon\to \xi}\ln \le(\frac{\Lambda(\epsilon)}{z_\xi(\epsilon)}\ri)\\
 \gamma_\eta&:=& \lim_{\epsilon\to \eta} \ln\le(\Lambda(\epsilon)z_\eta(\epsilon)\ri)\ .
 \eea
 Some remarks are in order: the function $\ln(\Lambda)$ is defined as any antiderivative of the 
 normalized third kind differential $\int_\xi^\eta\Omega$, which has residue $-1$ at $\xi$ and
 residue $+1$ at $\eta$. Hence $\Lambda$ has a simple zero at $\xi$ and a simple pole at $\eta$ 
 (in the simply connected domain). Also, $\Lambda$ is defined up to a multiplicative constant 
 depending on the base-point of integration: the final formula for the regularized integral does 
 not depend on this constant.
 $\Lambda$ can be written explicitly in terms of a Theta function and the $b$-periods of 
 $\frac{{\rm d}\Lambda}{\Lambda}$ are the difference of the Abel-map between $\xi$ and $\eta$.\par
 In the more general situation of the extended moduli space studied in Sect. \ref{extended} we
 had also some multivaluedness of the type
 \be
 \omega(\zeta+b_j)-\omega(\zeta)= {\rm d}H_j(\zeta)
 \ee
 where ${\rm d}H_j(\zeta)$, $j=1,\dots g$ are  meromorphic differential of the second kind with
 vanishing $a$-cycles. The formula for a regularized integral is easily adapted: the main 
 observation is that (\ref{genmaster}) now needs on the R.H.S. the following extra term
 \be
 \omega = (\hbox{\ref{genmaster}}) + \frac 1{2i\pi}\sum_{j=1}^g \oint_{a_j} H_j\Omega 
 \ee
 and consequently the formula for the regularized integral is 
 \be
 \slint_\xi^\eta\omega = (\hbox{\ref{slint}}) +  \frac 1{2i\pi}\sum_{j=1}^g 
 \oint_{a_j} H_j\frac{{\rm d}\Lambda}\Lambda
 \ee

\end{document}